\documentstyle[11pt,aaspp4]{article}
\def\la{\mathrel{\hbox{\rlap{\hbox{\lower4pt\hbox{$\sim$}}}\hbox{$<$}}}}
\def\ga{\mathrel{\hbox{\rlap{\hbox{\lower4pt\hbox{$\sim$}}}\hbox{$>$}}}}
\def\lesssim{\mathrel{\hbox{\rlap{\hbox{\lower4pt\hbox{$\sim$}}}\hbox{$<$}}}}
\def\etal{et al.\,\,}

\slugcomment{Accepted to {\it The Astrophysical Journal}}

\begin{document}

\title{Dust Extinction Curves and Ly$\alpha$ Forest Flux Deficits for Use in 
Modeling GRB Afterglows and All Other Extragalactic Point Sources}

\author{Daniel E. Reichart\altaffilmark{1,2,3}}

\altaffiltext{1}{Department of Astronomy and Astrophysics, University of
Chicago, 5640 South Ellis Avenue, Chicago, IL 60637}
\altaffiltext{2}{Department of Astronomy, California Institute of Technology,
Mail Code 105-24, 1201 East California Boulevard, Pasadena, CA 91125}
\altaffiltext{3}{Hubble Fellow}

\begin{abstract}

Since gamma-ray burst afterglows were first detected in 1997, the relativistic 
fireball model has emerged as the leading theoretical explanation of the 
afterglows.  In this paper, we present a very general, Bayesian inference 
formalism with which this, or any other, afterglow model can be tested, and with 
which the parameter values of acceptable models can be constrained, given the 
available photometry.  However, before model comparison or parameter estimation 
can be attempted, one must also consider the physical processes that affect the 
afterglow as it propagates along the line of sight from the burst source to the 
observer.  Namely, how does extinction by dust, both in the host galaxy and in 
our galaxy, and absorption by the Ly$\alpha$ forest and by H I in the host 
galaxy, change the intrinsic spectrum of the afterglow?  Consequently, we also 
present in this paper a very general, eight-parameter dust extinction curve 
model, and a two-parameter model of the Ly$\alpha$ forest flux deficit versus 
redshift distribution.  Using fitted extinction curves from Milky Way and 
Magellanic Cloud lines of sight, and measurements of Ly$\alpha$ forest flux 
deficits from quasar absorption line systems, we construct a Bayesian prior 
probability distribution that weights this additional, but necessary, parameter 
space such that the volume of the solution space is reduced significantly, {\it 
a priori}.  Finally, we discuss the broad applicability of these results to the 
modeling of light from all other extragalactic point sources, such as Type Ia 
supernovae.

\end{abstract}


\section{Introduction}

Optical afterglows have been detected for at least twelve gamma-ray bursts 
(GRBs); underlying galaxies have been detected for at least seven of these.  
Underlying galaxies have been detected by high-resolution imaging with HST [Sahu 
et al. 1997; Fruchter et al. 1999a (GRB 970228); Fruchter et al. 1999b (GRB 
970508); Kulkarni et al. 1998 (GRB 971214); Fruchter 1999b, private 
communication (GRB 980329); Bloom et al. 1999a; Fruchter et al. 1999c (GRB 
990123); Fruchter 1999c, private communication (GRB 990712)], by 
medium-resolution, ground-based imaging [e.g., Djorgovski et al. 1998a,b (GRB 
980613)], by detecting emission lines at afterglow locations [e.g., Djorgovski 
et al. 1998c (GRB 980613); Djorgovski et al. 1998d (980703)], and by sampling 
afterglow light curves until an asymptotic value is approached [e.g., Bloom et 
al. 1998 (GRB 980703)].  However, this last method is not always reliable, as 
Bloom et al. (1999b) have shown that a brightening supernova component to an 
afterglow light curve can be misinterpreted as being due to an underlying galaxy 
if the light curve is not sufficiently well-sampled at late times (see also 
Hjorth et al. 1999).
Lamb (1999) has shown that underlying galaxies that have been confirmed to be 
coincident with their afterglows by high-resolution, {\it HST} imaging are host 
galaxies to a high degree of certainty; however, $\approx 10 - 15$ \% of the 
remaining underlying galaxies are probably chance coincidences.  Consequently, 
at least six of these underlying galaxies are host galaxies, and the remaining 
one or two underlying galaxies are very likely to be host galaxies as well.

Since many, if not all, of the long bursts with detected optical afterglows are 
associated with host galaxies, these afterglows are likely to be extinguished by 
dust in their host galaxies (Reichart 1997), as well as by dust in our galaxy, 
and absorbed by H I in their host galaxies, as well as by the Ly$\alpha$ forest 
(Fruchter 1999a).  These physical processes affect - in some cases, probably 
significantly - the observed spectra of afterglows from the infrared (IR) 
through the ultraviolet (UV).  For example, Lamb \& Reichart (2000a,b) suggest 
that some of the $\approx 13$ bursts with securely-detected X-ray afterglows, 
but without securely-detected optical afterglows, might be explained by large 
amounts of extinction by dust in their host galaxies, probably from the 
immediate vicinities of these bursts if they are indeed associated with 
star-forming regions (see Lamb \& Reichart 2000a for a discussion of the 
evidence in favor of this association), or by absorption by the Ly$\alpha$ 
forest if these bursts occur at very high redshifts ($z \ga 5$).

Since the majority of afterglow observations are made at optical and 
near-infrared wavelengths, the effects of these physical processes on the 
observed spectra cannot be ignored, particularly since these spectra have been 
redshifted.  Indeed, these effects must be carefully modeled if intrinsic 
spectra are to be recovered.  This is the primary purpose of this paper.  The 
secondary purpose of this paper is to present a very general, Bayesian inference 
formalism with which afterglow models can be tested, and with which the 
parameter values of acceptable models can be constrained, given the available 
photometry.  We begin with this in \S 2.  Also in \S 2, we develop and present a 
formalism for the construction of Bayesian prior probability distributions from 
multi-dimensional data sets, which we draw on extensively in \S 4 and \S 5.  In 
\S 3, we present an eight-parameter dust extinction curve model, based on the 
work of Fitzpatrick \& Massa (1988) and Cardelli, Clayton, \& Mathis (1989).  In 
\S 4, we construct a prior that weights this additional parameter space such 
that the volume of the solution space is reduced significantly, {\it a priori}, 
using fitted extinction curves from Milky Way and Magellanic Cloud lines of 
sight.  In \S 5, we present a two-parameter model of the Ly$\alpha$ forest flux 
deficit versus redshift distribution, and we construct an analogous prior using 
Ly$\alpha$ forest flux deficit measurements from quasar absorption line systems. 
 In \S 6, we present a wide variety of extinguished and absorbed spectral flux 
distributions, using these models.  In \S 7, we draw conclusions, including a 
discussion of the broad applicability of these results to the modeling of light 
from all other extragalactic point sources.

\section{Statistical Methodology}

In \S 2.1, we present a very general, Bayesian inference formalism with which 
afterglow models (and any other model for that matter) can be tested, and with 
which the parameter values of acceptable models can be constrained, given the 
available photometry.  In \S 2.2, we develop and present a formalism for the 
construction of Bayesian prior probability distributions from multi-dimensional 
data sets, which we draw on extensively in \S 4 and \S 5.  For a deeper 
discussion of Bayesian inference, we refer the reader to an excellent review by 
Loredo (1992).

\subsection{Bayesian Inference}

\subsubsection{Bayes' Theorem}

Bayes' theorem states:
\begin{equation}
p(H|DI) = \frac{p(H|I)p(D|HI)}{p(D|I)},
\end{equation}
where $H$ is the hypothesis, or model, being considered, $D$ is the data, and 
$I$ is any available prior information.  Hence, Bayes' theorem states that the 
probability of a given hypothesis, $p(H|DI)$, given the data and any available 
prior information, is proportional to the product of the probability of the 
hypothesis, $p(H|I)$, given the prior information, and the probability of the 
data, $p(D|HI)$, given the hypothesis and the prior information.  The quantity 
$p(H|DI)$ is called the posterior probability distribution, the quantity 
$p(H|I)$ is called the prior probability distribution, and the quantity 
$p(D|HI)$ - sometimes denoted ${\cal L}(H)$ - is called the likelihood function.  

The quantity $p(D|I)$ normalizes the posterior.  Let the hypothesis, or model, 
$H$, be described by a set of parameters $\theta$.  Then, Bayes' theorem reads: 
\begin{equation}
p(\theta|DI) = \frac{p(\theta|I)p(D|\theta I)}{p(D|I)}.
\end{equation}
Normalization demands that
\begin{equation}
\int_{\theta} p(\theta|DI) d\theta = 1;
\label{norm}
\end{equation}
hence, $p(D|I)$ is given by
\begin{equation}
p(D|I) = \int_{\theta} p(\theta|I)p(D|\theta I)d\theta.
\label{post}
\end{equation}

Consequently, given a prior (see \S 2.1.2) and a likelihood function
(see \S 2.1.3), a normalized posterior may be computed.

\subsubsection{The Prior}

Let $\{\theta\}$ denote the region over which the parameters $\theta$ are 
integrated in Equations (\ref{norm}) and (\ref{post}).  The prior, 
$p(\theta|I)$, describes how any available pre-existing information constrains 
the values of the parameters $\theta$, or equivalently, how any available 
pre-existing information weights the parameter space $\{\theta\}$, and 
consequently, reduces the volume of the solution space, {\it a priori}.  

If no prior information is available, one usually takes the prior to be flat 
within a region $\{\theta_{phys}\} \subset \{\theta\}$ where the values of the 
parameters $\theta$ are considered to be physically plausible; the prior is 
taken to be zero everywhere else:
\begin{equation}
p(\theta|I) = \cases{(\int_{\theta_{phys}} d\theta)^{-1} & ($\theta \in 
\{\theta_{phys}\}$) \cr 0 & ($\theta \not\in \{\theta_{phys}\}$)};
\end{equation}
here, the volume integral normalizes the prior.\footnote{Here, we consider only 
linearly flat priors; however, logarithmically flat priors are also used, 
particularly when $\{\theta_{phys}\}$ spans orders of magnitude.}  The flat 
prior weights, {\it a priori}, all physically-plausible solutions equally, and 
gives no weight to physically-implausible solutions.

As an example, consider the case of a two-parameter model, where the parameters, 
$x$ and $y$, are physically unrelated.  In this case, the prior factorizes:
\begin{equation}
p(x,y|I) = p(x|I)p(y|I).
\end{equation}
Furthermore, suppose that prior information states that the possible values of 
$x$ are normally distributed with a mean of $a$ and a standard deviation of $b$, 
but that no prior information is available on the value of $y$, other than that 
values of $y < y_l$ and $y > y_u$ are considered physically implausible.  Then, 
the prior for the parameter $x$, given the prior information $a$ and $b$, is 
given by
\begin{equation}
p(x|a,b) = G(x,a,b), 
\end{equation}
where $G(x,a,b)$ is a normalized Gaussian distribution, given by
\begin{equation} 
G(x,a,b) = 
\frac{1}{\sqrt{2\pi}b}\exp\left[-\frac{1}{2}\left(\frac{x-a}{b}\right)^2\right],
\label{gauss}
\end{equation}
and the prior for the parameter $y$, given the prior information $y_l$ and 
$y_u$, is given by
\begin{equation}
p(y|y_l,y_u) = F(y,y_l,y_u),
\end{equation}
where $F(y,y_l,y_u)$ is a flat prior, given by
\begin{equation}
F(y,y_l,y_u) = \cases{(y_u-y_l)^{-1} & ($y_l < y < y_u$) \cr 0 & (otherwise)}.
\end{equation}

We make extensive use of Gaussian and flat priors in this paper, particularly in 
\S 2.2.  We present specific priors for parameters that describe the effects of 
extinction and absorption along the lines of sight to bursts in \S 4 and \S 5, 
respectively.

\subsubsection{The Likelihood Function}

The likelihood function, $p(D|\theta I)$, describes how any available data 
constrain the values of the parameters $\theta$, or equivalently, how any 
available data weight the parameter space $\{\theta\}$, and consequently, reduce 
the volume of the solution space.  Consequently, the posterior, $p(\theta|DI)$, 
which is proportional to the product of the prior and the likelihood function, 
describes how prior information and data jointly constrain the values of the 
parameters $\theta$, or equivalently, how prior information and data jointly 
weight the parameter space $\{\theta\}$, and consequently, jointly reduce the 
volume of the solution space.   

We now consider the form of the likelihood function for an unspecified afterglow 
model (and for any other spectral and temporal model for that matter).  Let 
$F_{\nu}(\nu,t;\theta)$ be the model's prediction for the spectral flux of an 
afterglow; $F_{\nu}(\nu,t;\theta)$ is a function of frequency of observation, 
$\nu$, time of observation, $t$, and the model parameters, $\theta$, which 
should include parameters that describe the effects of extinction and absorption 
along the line of sight.  Given $N$ measured spectral fluxes, the likelihood 
function is given by
\begin{equation}
p(D|\theta I) = \prod_{n=1}^N 
G_n[F_{\nu}(\nu_n,t_n;\theta),F_{\nu,n},\sigma_{F_{\nu},n}],
\end{equation}
where $F_{\nu,n}$ is the $n$th measured spectral flux, $\sigma_{F_{\nu},n}$ is 
the measured 1-$\sigma$ uncertainty associated with this spectral flux, and 
$G_n[F_{\nu}(\nu_n,t_n;\theta),F_{\nu,n},\sigma_{F_{\nu},n}]$ is a normalized 
Gaussian distribution, given by Equation (\ref{gauss}).

\subsubsection{Model Comparison}

Model comparison allows one to asses the relative probability of two or more 
models; consequently, this procedure may be used to reject non-viable models.  
Here, we consider the case of only two models; however, one can easily 
generalize the following procedure to the case of multiple models.

Consider two models, $H_{\theta}$ and $H_{\phi}$, that are described by two sets 
of parameters, $\theta$ and $\phi$, respectively.  The relative probability of 
model $H_{\theta}$ to model $H_{\phi}$ is called the odds ratio, and is given by
\begin{equation}
O_{\theta \phi} = 
\frac{\int_{\theta}p(\theta|DI)d\theta}{\int_{\phi}p(\phi|DI)d\phi} 
\end{equation}
\begin{equation}
= \frac{\int_{\theta}p(\theta|I)p(D|\theta 
I)d\theta}{\int_{\phi}p(\phi|I)p(D|\phi I)d\phi}.
\end{equation}
Normalization demands that
\begin{equation}
\int_{\theta} p(\theta|DI) d\theta + \int_{\phi} p(\phi|DI) d\phi = 1;
\end{equation} 
hence, given only models $H_{\theta}$ and $H_{\phi}$, the probability in favor 
of model $H_{\theta}$ is 
\begin{equation}
\int_{\theta} p(\theta|DI) d\theta = \frac{O_{\theta \phi}}{1+O_{\theta \phi}},
\end{equation}
and the probability in favor of model $H_{\phi}$ is 
\begin{equation}
\int_{\phi} p(\phi|DI) d\phi = \frac{1}{1+O_{\theta \phi}}.
\end{equation} 

\subsubsection{Parameter Estimation}

Parameter estimation allows one to constrain parameter values of acceptable 
models.  This procedure has two parts:  marginalization and the determination of 
credible regions.

Consider a single model, $H$, that is described by two sets of parameters:  
interesting parameters, $\theta$, whose values one wishes to constrain, and 
uninteresting parameters, $\phi$, whose values one does not need to constrain.  
Then, Bayes' theorem reads:
\begin{equation}
p(\theta\phi|DI) = \frac{p(\theta\phi|I)p(D|\theta\phi I)}{p(D|I)}.
\end{equation}
The posterior of the interesting parameters, $p(\theta|DI)$, is given by 
integrating the full posterior, $p(\theta\phi|DI)$, over the uninteresting 
parameters, $\phi$, and by then normalizing the resulting distribution:
\begin{equation}
p(\theta|DI) = \frac{\int_{\phi}p(\theta\phi|I)p(D|\theta\phi 
I)d\phi}{\int_{\theta\phi}p(\theta\phi|I)p(D|\theta\phi I)d\theta d\phi}.
\end{equation}
This procedure is called marginalization.  

Credible regions are determined by integrating the posterior from the most 
probable region of $\{\theta\}$ to the least probable region of $\{\theta\}$ 
until $p$ \% of the distribution has been integrated:
\begin{equation}
\int_{\theta_p} p(\theta|DI) d\theta = \frac{p}{100},
\end{equation}
where $\{\theta_p\} \subset \{\theta\}$ such that $p(\theta_1|DI) > 
p(\theta_2|DI)$ for any $\theta_1 \in \{\theta_p\}$ and for any $\theta_2 \in 
\{\theta\} - \{\theta_p\}$.  The region $\{\theta_p\}$ is called the $p$ \% 
credible region of the parameters $\theta$, or the solution space.  Of course, 
one can imagine many regions - actually an infinite number of regions - that 
integration over yields $p$ \% of the distribution; however, by integrating over 
the most probable region of $\{\theta\}$, one guarantees that the volume of 
$\{\theta_p\}$ is minimal, and that $\{\theta_p\}$ is uniquely defined.

\subsection{Constructing Priors from Multi-Dimensional Data Sets}

In \S 3, we present an eight-parameter model that describes the effects of 
extinction by dust along the lines of sight to bursts (and along the lines of 
sight to all extragalactic point sources for that matter).  However, without a 
prior that weights this parameter space such that the volume of the solution 
space is reduced significantly, {\it a priori}, this model has very little 
predictive power.  Fortunately, a considerable amount of prior information - in 
the form of fitted values for six of these eight parameters from 166 measured 
Milky Way and Magellanic Cloud extinction curves, and fitted values for one of 
the two remaining parameters from 79 of these extinction curves - exists; we 
describe this multi-dimensional data set in \S 4.  In this section, we develop 
and present a formalism by which priors can be extracted from multi-dimensional 
data sets; we draw on this formalism extensively in \S 4 and \S 5.  The 
extraction of a simply-formulated prior from, for example, the above, large, 
multi-dimensional data set greatly facilitates the incorporation of this prior 
information in future afterglow analyses.  We begin with a sequence of four 
illustrative examples in \S 2.2.1, the last three of which are particularly 
relevant to our construction of the dust extinction curve prior in \S 4.

\subsubsection{Examples in Three Dimensions and their Generalization}

With the first example, we describe the form that the prior should take in the 
ideal case of one (or more) of the quantities in the data set being fully 
determined by other quantities in the data set, and of this relation between 
these quantities being either known or easily determined from the data set.  
With the second example, we describe the form that the prior should take in the 
less ideal case of this relation between these quantities existing, but of its 
existence not being known or easily determined from the data set, although 
correlations between subsets of these quantities are determinable from the data 
set.  With the third example, we describe the form that the prior should take in 
the related case of this relation not being determinable from the data set 
because it involves quantities that are not in the data set, although 
correlations between the quantities that are in the data set, or subsets of 
these quantities, are determinable from the data set.  This last example is 
particularly realistic in that one often deals with physical processes, like 
dust extinction, that, although understood in general, many of the details of 
which depend on quantities whose relevance has not even been postulated yet, let 
alone whose values have been measured.  With the final example, we describe the 
form that the prior should take in the event that data selection effects, either 
due to instrumental limitations or due to how the sample was selected at a more 
human level, artificially constrain the values of quantities in the data set.  
We then discuss our generalization of these examples into a procedure.

\paragraph{Example 1.}
Consider a three-dimensional data set that consists of measured values of the 
parameters $w$, $x$, and $y$.  Furthermore, suppose that these parameters are 
related by $y = x + w$, and that $w$ and $x$ are physically-unrelated parameters 
whose measured values are distributed as the Gaussians $G(w,0,0.1)$ and 
$G(x,0,1)$, i.e., as $w = 0 \pm 0.1$ and $x = 0 \pm 1$.  Consequently, the 
measured values of the parameter $y$ should also be distributed as a Gaussian, 
namely, $G(y,0,1.005)$.  In this case, the prior that best represents this data 
set is given by
\begin{equation}
p(w,x,y|I) = G(w,0,0.1)G(x,0,1)\delta(y-x-w).
\end{equation}
This prior weights the three-dimensional parameter space, consequently reducing 
the volume of the solution space to a localized region of a two-dimensional 
plane, {\it a priori}.

\paragraph{Example 2.}  
Suppose now that the relation $y = x + w$ exists, but that its existence has not 
yet been determined.  One way to learn of this relation is to plot the data in 
three dimensions (or to plot the data in two dimensions and use perspective, or 
different symbols, or different colors, etc., to represent the third dimension). 
 However, this approach is increasingly difficult to implement in 
increasingly-higher dimensions.  Another approach is to probe the data set 
mathematically.  However, without prior knowledge of the form of the relation, 
let alone knowledge of its existence, this approach also can fail, particularly 
if the relation is non-linear in form.  Consequently, we now consider the form 
that the prior should take in the event that the relation $y = x + w$ is not 
known.  In this case, one approach is to plot the data two parameters at a time. 
 Having done this, one would immediately notice that the parameters $x$ and $y$ 
are strongly correlated, though it is unlikely that one would notice the weaker 
correlation between the parameters $w$ and $y$, since the values of the 
parameter $w$ span a much smaller range than do the values of the parameter $x$. 
 Finally, the parameters $w$ and $x$ also should appear to be uncorrelated, 
since we stated above that they are physically unrelated.  From this 
information, one can construct the following prior:
\begin{equation}
p(w,x,y|I) = G(w,0,0.1)G(x,0,1)G(y,x,0.1),
\label{prior}
\end{equation}
where the last two factors describe the distribution of the data in the $x$-$y$ 
plane.  Although this prior does not reduce the volume of the solution space as 
significantly as the above prior does, it certainly reduces it more than would 
the prior that one would construct if no correlations were noticed,
\begin{equation}
p(w,x,y|I) = G(w,0,0.1)G(x,0,1)G(y,0,1.005),
\end{equation}
and it certainly reduces the volume of the solution space significantly more 
than would the prior one would construct if the prior information were 
altogether ignored, i.e., if a flat prior were adopted (\S 2.1.2):
\begin{equation}
p(w,x,y|I) = F(w,w_l,w_u)F(x,x_l,x_u)F(y,y_l,y_u).
\end{equation}
Here, $w_l < w < w_u$, $x_l < x < x_u$, and $y_l < y < y_u$ define the ranges 
over which the values of these parameters are considered to be physically 
plausible. 

\paragraph{Example 3.}
Consider now the related case in which the parameter $w$ either is not or cannot 
be measured, and in fact, the very relevance of the parameter to the physical 
process at hand might not even be known.  In this case, the prior described by 
Equation (\ref{prior}) should be replaced by
\begin{equation}
p(w,x,y|I) = F(w,w_l,w_u)G(x,0,1)G(y,x,0.1),
\end{equation}
if $w$ is one of the model parameters, or by 
\begin{equation}
p(x,y|I) = G(x,0,1)G(y,x,0.1),
\end{equation}
if $w$ is not one of the model parameters.

\paragraph{Example 4.}
Finally, when constructing priors from data sets, one must be very careful that 
data selection effects do not bias the priors.  For example, suppose that the 
measured distribution of the parameter values of $x$ merely reflects how the 
data were sampled, and not how the parameter values of $x$ are intrinsically 
distributed.  In this case, the above prior should be replaced by 
\begin{equation}
p(x,y|I) = F(x,x_l,x_u)G(y,x,0.1).
\end{equation}
However, in this case, the factor $G(y,x,0.1)$ is an extrapolation beyond the 
range of the measured values of the parameter $x$, and must be treated as such.

When constructing a prior from a multi-dimensional data set in general, we adopt 
the following procedure:  (1) we plot two- and sometimes  three-dimensional 
subsets of the data to facilitate the identification of correlations between 
parameters; (2) if correlations are found, say between pairs of parameters, we 
determine the two-dimensional distributions that describe these subsets of the 
data; we also determine the one-dimensional distributions of the values of all 
of the parameters; and (3) we use this information to construct a prior for the 
full parameter space, as in the above examples, while being mindful of data 
selection effects.  How to go about steps (1) and (3) should be clear; how to go 
about step (2) - the construction of two-dimensional priors, as well as one- and 
three-dimensional priors - we explain in \S 2.2.2 and \S 2.2.3.

\subsubsection{Constructing Priors from Two-Dimensional Data Sets}

Suppose that two parameters, $x$ and $y$, are correlated, i.e., that the 
measured values of these parameters are scattered about a curve, $y = 
y_c(x;\theta_m)$, where $\theta_m$ are $M$ parameters that describe this curve.  
The scatter of these points about this curve can be both due to measurement 
errors, in which case the scatter is referred to as {\it intrinsic} scatter, and 
due to weaker dependences of either of the parameters $x$ or $y$ on other, 
yet-unmeasured, and even yet-unknown parameters (e.g., the parameter $w$ in 
Example 3 of \S 2.2.1), in which case the scatter is referred to as {\it 
extrinsic} scatter.  Below, we take all of these scatters to be normally 
distributed and uncorrelated.  Finally, let $g(x,y)\delta(y-y_c)$ be the 
intrinsic density of points along the curve $y = y_c(x;\theta_m)$, and let 
$f(x,y)$ be the selection function, i.e., the efficiency at which given values 
of the parameters $x$ and $y$ are observed.  We now construct a prior that 
describes the correlation between the parameters $x$ and $y$.

We model the intrinsic density of points in the $x'$-$y'$ plane by convolving 
the intrinsic density of points along the curve $y = y_c(x;\theta_m)$, i.e., 
$g(x,y)\delta(y-y_c)$, with the two-dimensional Gaussian smearing function 
$G(x',x,\sigma_x)G(y',y,\sigma_y)$, the scale of which is parameterized by 
1-$\sigma$ extrinsic scatters $\sigma_x$ and $\sigma_y$:
\begin{equation}
p_{int}(x',y'|\theta_m,\sigma_x,\sigma_y) = 
\int_x\int_yg(x,y)\delta(y-y_c)G(x',x,\sigma_x)G(y',y,\sigma_y)dxdy
\end{equation}
\begin{equation}
= \int_sg(x,y_c)G(x',x,\sigma_x)G(y',y_c,\sigma_y)ds,
\end{equation}
where $ds = \sqrt{dx^2+dy_c^2}$ is an element of path length along the length of 
the curve. 
The observed density of points in the $x'$-$y'$ plane is then given by
\begin{equation}
p_{obs}(x',y'|\theta_m,\sigma_x,\sigma_y) = 
f(x',y')p_{int}(x',y'|\theta_m,\sigma_x,\sigma_y)
\end{equation}
\begin{equation}
= \int_sf(x',y')g(x,y_c)G(x',x,\sigma_x)G(y',y_c,\sigma_y)ds.
\end{equation}

The probability distribution of the $n$th data point, $(x_n,y_n)$, given  
1-$\sigma$ intrinsic scatters $\sigma_{x,n}$ and $\sigma_{y,n}$, is given by  
\begin{equation}
p_n(x',y'|x_n,y_n,\sigma_{x,n}\sigma_{y,n}) = 
G_n(x',x_n,\sigma_{x,n})G_n(y',y_n,\sigma_{y,n}). 
\end{equation}
Hence, the joint probability distribution of a given model and the $n$th data 
point is given by 
\begin{equation}
p_n(x',y'|\theta_m,\sigma_x,\sigma_y,x_n,y_n,\sigma_{x,n}\sigma_{y,n}) = 
p_{obs}(x',y'|\theta_m,\sigma_x,\sigma_y)p_n(x',y'|x_n,y_n,\sigma_{x,n}\sigma_{y
,n})
\end{equation}
\begin{equation}
= 
\int_sf(x',y')g(x,y_c)G(x',x,\sigma_x)G(y',y_c,\sigma_y)G_n(x',x_n,\sigma_{x,n})
G_n(y',y_n,\sigma_{y,n})ds.
\end{equation}

The joint probability of a given model, i.e., given values of the parameters 
$\theta_m$, $\sigma_x$, and $\sigma_y$, and the $n$th data point is given by 
integrating 
$p_n(x',y'|\theta_m,\sigma_x,\sigma_y,x_n,y_n,\sigma_{x,n}\sigma_{y,n})$ over 
$x'$ and $y'$:
\begin{eqnarray}
\nonumber
p_n(\theta_m,\sigma_x,\sigma_y|x_n,y_n,\sigma_{x,n}\sigma_{y,n}) =
\end{eqnarray}
\begin{equation}  
\int_{x'}\int_{y'}\int_sf(x',y')g(x,y_c)G(x',x,\sigma_x)G(y',y_c,\sigma_y)G_n(x'
,x_n,\sigma_{x,n})G_n(y',y_n,\sigma_{y,n})dx'dy'ds.
\end{equation} 
Finally, the joint probability of a given model and {\it all} of the data points 
is given by taking the product of the $N$ probabilities 
$p_n(\theta_m,\sigma_x,\sigma_y|x_n,y_n,\sigma_{x,n}\sigma_{y,n})$:
\begin{eqnarray}
\nonumber
p(\theta_m,\sigma_x,\sigma_y|x_n,y_n,\sigma_{x,n},\sigma_{y,n}) =
\end{eqnarray}
\begin{equation} 
\prod_{n=1}^N\int_{x'}\int_{y'}\int_sf(x',y')g(x,y_c)G(x',x,\sigma_x) 
G(y',y_c,\sigma_y)G_n(x',x_n,\sigma_{x,n})G_n(y',y_n,\sigma_{y,n})dx'dy'ds.
\label{prior2}
\end{equation}
This is the prior.  In this form, it is a function of $M + 2$ parameters:  
$\theta_m$, $\sigma_x$, and $\sigma_y$.

If the scale over which the selection function $f(x',y')$ varies from constancy 
is larger than (1) the scale of the two-dimensional Gaussian 
$G(x',x,\sigma_x)G(y',y_c,\sigma_y)$, as measured by $\sigma_x$ and $\sigma_y$, 
and (2) the scale of the two-dimensional Gaussian 
$G_n(x',x_n,\sigma_{x,n})G_n(y',y_n,\sigma_{y,n})$, as measured by 
$\sigma_{x,n}$ and $\sigma_{y,n}$, then the first two integrations of Equation 
(\ref{prior2}) can be done analytically:
\begin{eqnarray}
\nonumber
p(\theta_m,\sigma_x,\sigma_y|x_n,y_n,\sigma_{x,n},\sigma_{y,n}) \approx
\end{eqnarray}
\begin{equation} 
\prod_{n=1}^Nf_n(x_n,y_n)\int_{x'}\int_{y'}\int_sg(x,y_c)G(x',x,\sigma_x)G(y',y_
c,\sigma_y)G_n(x',x_n,\sigma_{x,n})G_n(y',y_n,\sigma_{y,n})dx'dy'ds =
\end{equation}
\begin{equation} 
\prod_{n=1}^Nf(x_n,y_n)\int_{s}g(x,y_c)G_n\left(x,x_n,\sqrt{\sigma_x^2+\sigma_{x
,n}^2}\right)G_n\left(y_c,y_n,\sqrt{\sigma_y^2+\sigma_{y,n}^2}\right)ds.
\label{prior3}
\end{equation}

The final integration, however, is non-trivial.  It consists of a path 
integration through the product of two distributions:  $g(x,y)\delta(y-y_c)$, 
the intrinsic density of points along the curve $y = y(x;\theta_m)$, and the 
two-dimensional Gaussian $G_n\left(x,x_n,\sqrt{\sigma_x^2+\sigma_{x,n}^2}\right) 
G_n\left(y,y_n,\sqrt{\sigma_y^2+\sigma_{y,n}^2}\right)$.  However, if the scale 
of this two-dimensional Gaussian, as measured by  
$\sqrt{\sigma_x^2+\sigma_{x,n}^2}$ and $\sqrt{\sigma_y^2+\sigma_{y,n}^2}$, is 
smaller than (1) the scale over which $y_c(x;\theta_m)$ varies from linearity, 
and (2) the scale over which $g(x,y_c)$ varies from constancy, this integration 
can be done with relative ease, as we now show.

Let $(x_{t,n},y_{t,n})$ be the point on the curve $y = y_c(x;\theta_m)$ for 
which the value of the two-dimensional Gaussian, 
$G_n\left(x,x_n,\sqrt{\sigma_x^2+\sigma_{x,n}^2}\right) 
G_n\left(y,y_n,\sqrt{\sigma_y^2+\sigma_{y,n}^2}\right)$, is maximum.  At this 
point, the curve $y = y_c(x;\theta_m)$ will be tangent to an iso-contour of the 
two-dimensional Gaussian, i.e., the ellipse given by
\begin{equation}
\frac{(x-x_n)^2}{\sigma_x^2+\sigma_{x,n}^2} + 
\frac{(y-y_n)^2}{\sigma_y^2+\sigma_{y,n}^2} = 
\frac{(x_{t,n}-x_n)^2}{\sigma_x^2+\sigma_{x,n}^2} + 
\frac{(y_{t,n}-y_n)^2}{\sigma_y^2+\sigma_{y,n}^2}.
\end{equation}
By repeatedly setting the slope of this tangential ellipse equal to the slope of 
the curve, the tangent point, $(x_{t,n},y_{t,n})$, can be found iteratively; if 
$y_c(x;\theta_m)$ is indeed slowly varying, only a few iterations are required.  
Now, making using of the first assumption - that $y_c(x;\theta_m)$ does not vary 
significantly from linearity over the scale of the two-dimensional Gaussian - 
one can replace $y_c(x;\theta_m)$ in Equation (\ref{prior3}) with the following 
approximation:
\begin{equation}
y_c(x;\theta_m) \approx y_{t,n} + s_{t,n}(x-x_{t,n}),
\end{equation}
where 
\begin{equation}
s_{t,n} = \left[\frac{\partial y_c(x;\theta_m)}{\partial x}\right]_{x=x_{t,n}}
\end{equation}
is the slope of $y_c(x;\theta_m)$ (or that of the tangential ellipse) at the 
tangent point $(x_{t,n},y_{t,n})$.  Finally, by making use of the second 
assumption - that $g(x,y_c)$ does not vary significantly from constancy over the 
scale of the two-dimensional Gaussian - one can complete the integration of 
Equation (\ref{prior3}) analytically:
\begin{eqnarray}
\nonumber
p(\theta_m,\sigma_x,\sigma_y|x_n,y_n,\sigma_{x,n},\sigma_{y,n}) \approx 
\end{eqnarray}
\begin{equation} 
\prod_{n=1}^Nf_n(x_n,y_n)g_n(x_n,y_n)\int_{s} 
G_n\left(x,x_n,\sqrt{\sigma_x^2+\sigma_{x,n}^2}\right) 
G_n\left(y_c,y_n,\sqrt{\sigma_y^2+\sigma_{y,n}^2}\right)ds \approx
\end{equation}
\begin{equation} 
\prod_{n=1}^Nf_n(x_n,y_n)g_n(x_n,y_n)\sqrt{1+s_{t,n}^2}
G_n\left[y_n,y_{t,n}+s_{t,n}(x_n-x_{t,n}),\sqrt{\sigma_y^2+\sigma_{y,n}^2+s_{t,n
}^2(\sigma_x^2+\sigma_{x,n}^2)}\right].
\end{equation}
Normalization of the prior removes the need to determine the value of 
the constant $\prod_{n=1}^Nf_n(x_n,y_n)g_n(x_n,y_n)$; hence,
\begin{eqnarray}
\nonumber
p(\theta_m,\sigma_x,\sigma_y|x_n,y_n,\sigma_{x,n},\sigma_{y,n}) \approx 
\end{eqnarray}
\begin{equation} 
\prod_{n=1}^N\sqrt{1+s_{t,n}^2} 
G_n\left[y_n,y_{t,n}+s_{t,n}(x_n-x_{t,n}),\sqrt{\sigma_y^2+\sigma_{y,n}^2+s_{t,n
}^2(\sigma_x^2+\sigma_{x,n}^2)}\right].
\label{prior4}
\end{equation}

\subsubsection{Constructing Practical Priors from Two-Dimensional Data Sets and 
its Generalization}

From the point of view of practicality, Equation (\ref{prior4}) has a
number of drawbacks.  First of all, by formulating the prior in this
way, we have replaced the two parameters $x$ and $y$ with $M + 2$,
intermediate parameters:  $\theta_m$, $\sigma_x$, and $\sigma_y$. 
Secondly, potential users of this prior must have access to the $4N$ pieces of 
prior information, $x_n$, $y_n$, $\sigma_{x,n}$, and $\sigma_{y,n}$, where $N$ 
can be a very large number, that are required for its computation.  Finally, the 
computation of this prior, although completely feasible, is non-trivial:  the 
iterative procedure of finding
the tangent point (\S 2.2.2) must be performed $N$ times at every grid point in 
the $(M+2)$-dimensional space that the prior spans.

These problems can be overcome by instead taking Equation
(\ref{prior4}) to be a likelihood function, and by then applying the
statistical methodology of \S 2.1.5 to constrain the values of the
intermediate parameters $\theta_m$, $\sigma_x$, and $\sigma_y$, i.e.,
to reduce the $4N$ pieces of prior information to what we show below to
be $2M + 2$ representative values, where $M$ is typically a few.  Given
these fitted values, it is a simple matter to construct an
approximation to Equation (\ref{prior4}) (1) that is solely a function
of the parameters $x$ and $y$, (2) that requires only these $2M + 2$ values as 
prior information, and (3) that is computationally
non-taxing.  We do this now; we then generalize these results to other 
dimensions.

Let $\hat{\theta}_m$, $\hat{\sigma}_x$, and $\hat{\sigma}_y$ be the best-fit 
values of $\theta_m$, $\sigma_x$, and $\sigma_y$, and let 
$\hat{\sigma}_{\theta_m}$ be the fitted, 1-$\sigma$ uncertainties in the values 
of $\hat{\theta}_m$.  If one takes these fitted values to be normally 
distributed and uncorrelated, Equation (\ref{prior4}) may then be approximated 
by
\begin{equation}
p(x,y|\hat{\theta}_m,\hat{\sigma}_{\theta_m},\hat{\sigma}_x,\hat{\sigma}_y) 
\approx 
G[y,y_c(x;\hat{\theta}_m),\sigma_y(x;\hat{\theta}_m,\hat{\sigma}_{\theta_m},\hat
{\sigma}_x,\hat{\sigma}_y)], 
\label{prior5}
\end{equation}
where
\begin{equation}
\sigma_y(x;\hat{\theta}_m,\hat{\sigma}_{\theta_m},\hat{\sigma}_x,\hat{\sigma}_y) 
= \left\{\sum_{m=1}^M\left[\sigma_{\theta_m}\frac{\partial 
y_c(x;\theta_m)}{\partial \theta_m}\right]_{\theta_m=\hat{\theta}_m}^2 + 
\left[\sigma_x\frac{\partial y_c(x;\theta_m)}{\partial 
x}\right]_{\theta_m=\hat{\theta}_m}^2 + \sigma_y^2\right\}^{1/2}.
\label{sigma}
\end{equation}
Here, 
$\sigma_y(x;\hat{\theta}_m,\hat{\sigma}_{\theta_m}\hat{\sigma}_x,\hat{\sigma}_y)
$ is the quadratic sum of the uncertainty in the curve $y = 
y_c(x;\hat{\theta}_m)$ due to the uncertainties $\hat{\sigma}_{\theta_m}$ in the 
best-fit values $\hat{\theta}_m$, the uncertainty in the curve $y = 
y_c(x;\hat{\theta}_m)$ due to the extrinsic scatter $\hat{\sigma}_x$ in the $x$ 
dimension, and the uncertainty in the curve $y = y_c(x;\hat{\theta}_m)$ due to 
the extrinsic scatter $\hat{\sigma}_y$ in the $y$ dimension.  
Consequently, Equation (\ref{prior5}) (1) is indeed solely a function of the 
parameters $x$ and $y$, (2) requires only $2M+2$ pieces of prior information, 
$\hat{\theta}_m$, $\hat{\sigma}_{\theta_m}$, $\hat{\sigma}_x$, and 
$\hat{\sigma}_y$, and (3) is easily computed.

Equation (\ref{prior5}) can be improved upon if the selection function, 
$f(x,y)$, is well understood.  In this case, the intrinsic density of points 
along the curve $y = y_c(x;\theta_m)$, i.e., $g(x,y_c)$, can be recovered from 
the observed density of points along the curve $y = y_c(x;\theta_m)$, i.e., 
$f(x,y_c)g(x,y_c)$, in which case, Equation (\ref{prior5}) can be replaced with 
\begin{equation}
p(x,y|\hat{\theta}_m,\hat{\sigma}_{\theta_m},\hat{\sigma}_x,\hat{\sigma}_y) 
\approx g[x,y_c(x;\hat{\theta}_m)] 
G[y,y_c(x;\hat{\theta}_m),\sigma_y(x;\hat{\theta}_m,\hat{\sigma}_{\theta_m},\hat
{\sigma}_x,\hat{\sigma}_y)]. 
\end{equation}
However, selection functions often are not well understood, as is the case with 
the data sets that we present in \S 4 and \S 5; consequently, we do not develop 
this case further in this paper.

One-dimensional priors are trivially derived by setting $s_{t,n} = 0$ in 
Equation (\ref{prior4}).  In this case, Equation (\ref{prior5}) reduces to:
\begin{equation}
p(y|\hat{y},\hat{\sigma}_y) = G(y,\hat{y},\hat{\sigma}_y).
\end{equation}

Furthermore, it is not difficult to generalize Equations (\ref{prior4}) and 
(\ref{prior5}) to more than two dimensions.  Let $y = y_c(x_l|\theta_m)$, where 
$1 \le l \le L$.  In this case, one can show that Equation (\ref{prior4}) 
generalizes to
\begin{eqnarray}
\nonumber
p(\theta_m,\sigma_{x_l},\sigma_y|x_{l,n},y_n,\sigma_{x_l,n},\sigma_{y,n}) 
\approx 
\end{eqnarray}
\begin{equation} 
\prod_{n=1}^N\sqrt{1+\sum_{l=1}^Ls_{l,t,n}^2} 
G_n\left[y_n,y_{t,n}+\sum_{l=1}^Ls_{l,t,n}(x_{l,n}-x_{l,t,n}), 
\sqrt{\sigma_y^2+\sigma_{y,n}^2+\sum_{l=1}^Ls_{l,t,n}^2(\sigma_{x_l}^2+\sigma_{x
_l,n}^2)}\right],
\end{equation}
where
\begin{equation}
s_{l,t,n} = \left[\frac{\partial y_c(x_l;\theta_m)}{\partial 
x_l}\right]_{x_l=x_{l,t,n}}, 
\end{equation}
and Equation (\ref{prior5}) generalizes to
\begin{equation}
p(x_l,y|\hat{\theta}_m,\hat{\sigma}_{x_l},\hat{\sigma}_y) \approx 
G[y,y_c(x_l;\hat{\theta}_m),\sigma_y(x_l;\hat{\theta}_m,\hat{\sigma}_{\theta_m},
\hat{\sigma}_{x_l},\hat{\sigma}_y)], 
\end{equation}
where
\begin{equation}
\sigma_y(x_l;\hat{\theta}_m,\hat{\sigma}_{\theta_m} 
\hat{\sigma}_{x_l},\hat{\sigma}_y) = 
\left\{\sum_{m=1}^M\left[\sigma_{\theta_m}\frac{\partial 
y_c(x_l;\theta_m)}{\partial \theta_m}\right]_{\theta_m=\hat{\theta}_m}^2 + 
\sum_{l=1}^L\left[\sigma_{x_l}\frac{\partial y_c(x_l;\theta_m)}{\partial 
x_l}\right]_{\theta_m=\hat{\theta}_m}^2 + \sigma_y^2\right\}^{1/2}.
\end{equation}

\section{The Dust Extinction Curve Model}

We now present an eight-parameter model that describes the effects of extinction 
by dust along lines of sight through our galaxy, and by redshifting this model, 
along lines of sight through burst host galaxies (and along lines of sight 
through the host galaxies of all extragalactic point sources for that matter).  
This model is a combination of the two-parameter, IR and optical extinction 
curve model of Cardelli, Clayton, \& Mathis (1989), and the eight-parameter, UV 
extinction curve model of Fitzpatrick \& Massa (1988).  We present the IR and 
optical extinction curve model in \S 3.1; we present the UV extinction curve 
model in \S 3.2.  In \S 3.3, we modify these models to include the effect at 
far-UV (FUV) wavelengths of absorption by H I in galaxies.

\subsection{$\lambda > 3000$ ${\rm\AA}$}

Using UBVRIJHKL photometry of 29 reddened Milky Way OB stars (Clayton \& Mathis 
1988; Clayton \& Cardelli 1988), and UV extinction curves that had been fitted 
to International Ultraviolet Explorer (IUE) spectra of 45 Milky Way OB stars 
(Fitzpatrick \& Massa 1988; see \S 3.2), Cardelli, Clayton, \& Mathis (1988, 
1989) constructed an empirical, two-parameter, IR through FUV extinction curve 
model.  The two parameters are $A_V$ and $R_V$.  The former parameter normalizes 
the extinction curve at the V band; the latter parameter, defined by
\begin{equation}
R_V = \frac{A_V}{E(B-V)}
\label{rv}
\end{equation}
\begin{equation}
= \frac{1}{\frac{A_B}{A_V}-1},
\label{rv2}
\end{equation}
is a measure of the amount of extinction at the B band relative to that at the V 
band.  The standard diffuse interstellar medium (ISM) value of $R_V$ is 3.1; 
however, the value of $R_V$ is known to vary with the type of interstellar 
environment.  For example, $R_V \sim 4 - 5$ is typical of dense clouds.  

The Cardelli, Clayton, \& Mathis (1989) extinction curve is given by:
\begin{equation}
\frac{A_\lambda}{A_V} = a(x) + \frac{b(x)}{R_V},
\label{ccm}
\end{equation}
where $x = (\lambda / 1$ $\mu m)^{-1}$, and $a(x)$ and $b(x)$ are empirical 
expressions given by
\begin{equation}
a(x) = \cases{0.574x^{1.61} & ($0.3 < x < 1.1$) \cr 1 + 0.17699y - 0.50447y^2 - 
0.02427y^3 + 0.72085y^4 \cr + 0.01979y^5 - 0.77530y^6 + 0.32999y^7 & ($1.1 < x < 
3.3$)}
\end{equation}
and
\begin{equation}
b(x) = \cases{-0.527x^{1.61} & ($0.3 < x < 1.1$) \cr 1.41338y + 2.28305y^2 + 
1.07233y^3 - 5.38434y^4 \cr - 0.62251y^5 + 5.30260y^6 - 2.09002y^7 & ($1.1 < x < 
3.3$)}, 
\end{equation}
where $y = x - 1.82$.
Cardelli, Clayton, and Mathis (1989) also determined expressions for $a(x)$ and 
$b(x)$ in the wavelength range $3.3 < x < 10$ (1000 \AA$ $ $< \lambda < 3000$ 
\AA); however, the Fitzpatrick \& Massa (1988) parameterization of the 
extinction curve, on which this portion of the Cardelli, Clayton, \& Mathis 
(1989) parameterization of the extinction curve is largely based, is a more 
general description of the extinction curve in the UV.  Consequently, we instead 
adopt the more general extinction curve model of Fitzpatrick \& Massa (1988) at 
these UV wavelengths (see \S 3.2).

The extinction curve at wavelengths $\lambda \ga 6000$ \AA$ $ is generally 
attributed to absorption and scattering by classical Van de Hulst grains (Van de 
Hulst 1957).  These are relatively large grains, with sizes of 1000 -- 2000 \AA. 
 They are thought to be fluffy, non-spherical composites containing carbon, 
silicates, oxides, and vacuum (Mathis 1996, 1998; Dwek 1998).  Classical grain 
extinction saturates at a wavelength of $\lambda \sim 3000$ \AA.

\subsection{$1000$ ${\rm\AA} < \lambda < 3000$ ${\rm\AA}$}

In a series of four papers, Massa \& Fitzpatrick (1986) and Fitzpatrick \& Massa 
(1986, 1988, 1990) measured extinction curves for two samples of reddened Milky 
Way OB stars from IUE spectra.  
Their cluster sample consists of 35 stars from five clusters; since these stars 
were drawn from similar interstellar environments, their extinction curves are 
relatively similar.  Their program sample consists of 45 stars from a wide 
variety of interstellar environments; consequently, the extinction curves of 
this sample are more varied.  
Fitzpatrick \& Massa (1988, 1990) found that all 80 extinction curves are well 
fitted by the following three-component function:
\begin{equation}
\frac{E(\lambda-V)}{E(B-V)} = c_1 + c_2x + c_3D(x;\gamma,x_0) + c_4F(x),
\label{fm}
\end{equation}
where 
\begin{equation}
D(x;\gamma,x_0) = \frac{x^2}{(x^2-x_0^2)^2+x^2\gamma^2}
\label{drude}
\end{equation}
and
\begin{equation}
F(x) = \cases{0.5392(x-5.9)^2 + 0.05644(x-5.9)^3 & ($x > 5.9$) \cr 0 & ($x < 
5.9$)}.
\end{equation}
The first component, $c_1 + c_2x$, is linear and spans the wavelength range of 
the data:  $1000$ \AA$ $ $< \lambda < 3000$ \AA.  The second component, 
$D(x;\gamma,x_0)$, is a functional form called the Drude profile; however, in 
this context, it is often called the 2175 \AA$ $ bump, or the UV bump.  The 
third component, $F(x)$, is an empirical expression called the FUV curvature 
component, or the FUV non-linear component.  We depict all three of these 
components in Figure 1.

Although the Fitzpatrick \& Massa (1988) parameterization of the extinction 
curve is first and foremost an empirically-driven fitting function, it is not 
devoid of physical significance.  For example, the Drude profile is the 
functional form of the absorption cross section of a forced-damped harmonic 
oscillator; it reduces to a Lorentzian near resonance (Jackson 1962).  
Fitzpatrick \& Massa (1986) found that the Drude profile better fits the data 
than does a pure Lorentzian;   Lorentzian profiles had been used previously 
(Savage 1975; Seaton 1979).  The Drude profile is a function of $x_0$, the 
bump's center, and $\gamma$, the bump's width.

The linear component of the extinction curve is generally attributed to a 
distribution of grain sizes; the larger grains, with sizes perhaps as large as 
the classical grains, are responsible for extinction in the near-UV, and the 
smaller grains, with sizes perhaps as small as 100 \AA$ $ or less, are 
responsible for extinction in the FUV.  These grains have been interpreted 
either as the tail end of the classical grain population (e.g., Mathis, Rumpl, 
\& Nordsieck 1977), or as a separate population altogether (e.g., Hong \& 
Greenberg 1980).  The parameters $c_1$ and $c_2$ are correlated (see Figure 2, 
\S 4.2);  however, this correlation is merely an artifact of the fitting 
procedure by which the values of these parameters are determined (Carnochan 
1986, see \S 4.2).

The values of these parameters are known to vary with the type of interstellar 
environment.  For example, in the diffuse ISM, the values of $c_2$ are in the 
range 0.6 -- 1, while in dense clouds, the values of $c_2$ extend to lower 
values: 0 -- 1 (e.g., Fitzpatrick \& Massa 1988).  This difference is generally 
attributed to the accretion of small grains onto larger grains, or to the 
coagulation of small grains into larger grains, both of which occur most readily 
in dense clouds (Scalo 1977; Cardelli, Clayton, \& Mathis 1988, 1989; Mathis \& 
Wiffen 1989).
In young star-forming regions, like the Orion Nebula, which are also dense 
clouds, $c_2 \approx 0$ (e.g., Fitzpatrick \& Massa 1988).  This consistent, low 
value is generally attributed to stellar radiation forces, or to the evaporation 
of grains, both of which preferentially remove the smaller grains (McCall 1981; 
Cardelli \& Clayton 1988).  

Dense clouds also protect grains from supernovae shocks, which preferentially 
destroy the larger, classical grains or their mantles, and thus possibly 
increase the number of grains responsible for the linear component of the 
extinction curve (Seab \& Shull 1983; Jenniskens \& Greenberg 1993; Jones, 
Tielens, \& Hollenbach 1996).  Since extinction curves are normalized at the V 
band, the removal of classical grains alone guarantees higher values of $c_2$ 
(e.g., Jenniskens \& Greenberg 1993).  
Indeed, in the Large and Small Magellanic Clouds (LMC and SMC), where old 
star-forming regions, like 30 Doradus (an extreme example), are more common than 
in the Galaxy, the values of $c_2$ extend to higher values:  0.6 -- 2 for the 
LMC and the SMC wing, and 2 -- 2.5 for the SMC bar (Calzetti, Kinney, \& 
Storchi-Bergmann 1994; Gordon \& Clayton 1998; Misselt, Clayton, \& Gordon 
1999).  In starburst galaxies, the values of $c_2$ are similarly high (Gordon, 
Calzetti, \& Witt 1997).  

There is less consensus about the type of grain that is responsible for the UV 
bump; however, the fact that for fixed values of the IR and optical extinction 
curve parameter $R_V$, and of the linear component parameters $c_1$ and $c_2$, 
the strength of the UV bump can vary considerably, strongly suggests that the 
classical and linear component grains are not responsible for the UV bump (e.g., 
Greenberg \& Chlewicki 1983).  
The UV bump is sometimes attributed to small graphitic grains, having diameters 
of $\sim 200$ \AA$ $ or less (e.g., Hecht 1986).  One property of this model is 
that the bump's width, $\gamma$, can vary by a few tens of percent, while its 
center, $x_0$, can vary by only a few percent; this is what is observed (e.g., 
Fitzpatrick \& Massa 1986).  Another property of this model is that $c_3$ and 
$\gamma$ are correlated, which is also observed (see Figure 4, \S 4.2); however, 
this correlation appears to change with the type of interstellar environment 
(see \S 4.2).  

Equations (\ref{fm}) and (\ref{drude}) show that the height, or strength, of the 
UV bump is proportional to $c_3 / \gamma^2$.  Dense clouds and the diffuse ISM 
tend to favor strong UV bumps; however, star-forming regions, both young and 
old, tend to favor weak UV bumps.  In fact, in the SMC bar and starburst 
galaxies, no UV bump is typically observed.  Young star-forming regions tend to 
favor weak UV bumps probably because UV radiation destroys or alters the grains 
that are responsible for the UV bump (Jenniskens \& Greenberg 1993); old 
star-forming regions tend to favor weak UV bumps probably because UV radiation 
and/or supernova shocks destroy or alter the grains that are responsible for the 
UV bump (Gordon, Calzetti, \& Witt 1997).  These effects can be seen in Figure 5 
(see \S 4.2; see also Clayton, Gordon, \& Wolff 2000), where $c_3$ is an 
approximate measure of the strength of the UV 
bump.

Even less is known about the type of grain that is responsible for the FUV 
non-linear component of the extinction curve.  According to the model of Hecht 
(1986), the small graphitic grains that produce the UV bump should have a second 
plasmon resonance, resulting in a second and similar bump centered at a 
wavelength of $\lambda \sim 700 - 800$ \AA.  Indeed, the shape of the FUV 
non-linear component resembles the red wing of a Drude profile (Fitzpatrick \& 
Massa 1988).  
Furthermore, this model suggests that the strengths of these two bumps might be 
correlated (e.g., Fitzpatrick \& Massa 1988); however, hydrogenation of these 
grains should largely decouple these resonances (e.g., Hecht 1986).  Fitzpatrick 
\& Massa (1988) and Jenniskens \& Greenberg (1993) have shown that $c_3$ and 
$\gamma$ are both weakly correlated with $c_4$; however, inclusion of the LMC 
and SMC lines of sight suggests that these correlations also depend on the type 
of interstellar environment (see Figure 6, \S 4.2).  As in the case of the UV 
bump parameters, $c_4$ does not correlate with $R_V$, $c_1$, or $c_2$ 
(Jenniskens \& Greenberg 1993).

Like the IR and optical extinction curve model of Cardelli, Clayton, \& Mathis 
(1989; Equation \ref{ccm}), the UV extinction curve model of Fitzpatrick \& 
Massa (1988; Equation \ref{fm}) can be written as a function of 
$A_{\lambda}/A_V$, given the following rearrangement of the definition of $R_V$ 
(Equation \ref{rv}):
\begin{equation}
\frac{A_\lambda}{A_V} = 1 + \frac{1}{R_V}\frac{E(\lambda-V)}{E(B-V)}.
\end{equation}
To smoothly link these two models (see also Fitzpatrick 1999), we recommend the 
following weighted average 
of these models between the V band ($\lambda = 5500$ \AA) and $\lambda = 3000$ 
\AA:
\begin{equation}
A_\lambda = \cases{A_{\lambda,CCM} & ($x < 1.82$) \cr A_{\lambda,CCM} + 
\frac{x-1.82}{1.48}(A_{\lambda,FM}-A_{\lambda,CCM}) & ($1.82 < x < 3.3$) \cr 
A_{\lambda,FM} & ($x > 3.3$)}.
\label{ext}
\end{equation}

To summarize, to the limit of current observations, the IR through FUV 
extinction curve appears to be most generally modeled by eight parameters:  
$A_V$, $R_V$, $c_1$, $c_2$, $c_3/\gamma^2$, $c_4$, $\gamma$, and $x_0$; the UV 
bump is more naturally parameterized by $c_3/\gamma^2$ (bump height) and 
$\gamma$ (bump width) than by $c_3$ and $\gamma$, since the former pair is 
orthogonal (Jenniskens \& Greenberg 1993).
In \S 3.3, we modify Equation (\ref{ext}) to include the effect at FUV 
wavelengths of absorption by H I in galaxies; this does not change the number of 
parameters.  In \S 4, we show that the volume of the solution space of this 
additional, but necessary, eight-dimensional parameter space can be reduced 
significantly, {\it a priori}, with a prior; without such a prior, Equation 
(\ref{ext}) has very little predictive power.

\subsection{$\lambda < 1000$ ${\rm\AA}$}

The column density of H I in a galaxy along a line of sight is given by
\begin{equation}
N_H = E(B-V)\eta
\end{equation} 
\begin{equation}
= 1.5 \times 10^{21}\,{\rm cm}^{-2}\,\left(\frac{A_V}{1\,{\rm 
mag}}\right)\left(\frac{R_V}{3.1}\right)^{-1}\left(\frac{\eta}{\eta_{MW}}\right)
, 
\end{equation}
where $\eta$ is the gas to dust ratio of the galaxy along the line of sight, and 
$\eta_{MW}$ is the standard value of this ratio for the Milky Way.  The 
bound-free photo-absorption cross section of ground state hydrogen as a function 
of wavelength is given by  
\begin{equation}
\alpha_\lambda = \cases{7.9 \times 10^{-18}\,{\rm 
cm}^2\,\left(\frac{\lambda}{912\,\AA}\right)^3 & ($\lambda < 912$ \AA) \cr 
0\,{\rm cm}^2 & ($\lambda > 912$ \AA)}.
\end{equation}
Total extinction occurs when $\alpha_\lambda N_H \gg 1$, i.e., when
\begin{equation}
A_V \gg 8.2 \times 
10^{-5}\,\left(\frac{R_V}{3.1}\right)\left(\frac{\eta}{\eta_{MW}}\right)^{-1} 
\left(\frac{\lambda}{912\,\,{\rm\AA}}\right)^{-3}\,{\rm mag}
\end{equation}
and $\lambda < 912$ \AA.  Since this condition is always satisfied along lines 
of sight through galaxies, at least into the soft X rays, we replace Equation 
(\ref{ext}) with:
\begin{equation}
A_\lambda = \cases{A_{\lambda,CCM} & ($x < 1.82$) \cr A_{\lambda,CCM} + 
\frac{x-1.82}{1.48}(A_{\lambda,FM}-A_{\lambda,CCM}) & ($1.82 < x < 3.3$) \cr 
A_{\lambda,FM} & ($3.3 < x < 10.96$) \cr \infty & ($x > 10.96$)}.
\label{ext2}
\end{equation}
Consequently, Equation (\ref{ext2}) describes how light is extinguished by dust 
and absorbed by H I along lines of sight through galaxies, in the rest frame, 
into the soft X rays.

\section{The Dust Extinction Curve Prior}

Using fitted extinction curves from Milky Way and Magellanic Cloud lines of 
sight, and the statistical methodology that we presented in \S 2, we now 
construct a prior that weights the eight-dimensional parameter space of the 
extinction curve model that we presented in \S 3 such that the volume of the 
solution space is reduced significantly, {\it a priori}.  We describe the data 
set in \S 4.1.  We model correlations between extinction curve parameters, and 
construct the prior, in \S 4.2.

\subsection{The Data Set}

From the literature, we have collected the results of fits to 166 extinction 
curves:  we know the values of the UV extinction curve parameters $c_1$, $c_2$, 
$c_3$, $c_4$, $\gamma$, and $x_0$ for all of these lines of sight from fits to 
IUE spectra; we know the values of the IR and optical extinction curve parameter 
$R_V$ for 79 of these lines of sight from IR and optical photometry.  
These lines of sight sample a wide variety of interstellar environments in the 
Milky Way, the LMC, and the SMC.  We describe the breakdown of the data set in 
detail below; we summarize this information in Table 1.

Ideally, we would fit the models of correlations between extinction curve 
parameters that we present in \S 4.2 directly to the IUE spectra and IR and 
optical photometry, instead of to the fitted values of the extinction curve 
parameters, which represent a compression of the information contained in the 
actual data.  However, to do so would not be practicable.  Consequently, we 
instead adopt the best-fit values of the extinction curve parameters, $c_1$, 
$c_2$, $c_3$, $c_4$, $\gamma$, $x_0$, and $R_V$, and the fitted uncertainties in 
these parameters, when available, as the data set, at the expense of a minor 
loss of information.

We have drawn the results of fits to Galactic extinction curves from two 
sources:  the cluster and program samples of Fitzpatrick \& Massa (1990) and the 
sample of Jenniskens \& Greenberg (1993).  The FM cluster sample consists of 35 
fitted extinction curves, the FM program sample consists of 45 fitted extinction 
curves, and the JG sample consists of 115 fitted extinction curves.  For 
information about how these samples were selected, and about how the selected, 
IUE spectra were fitted, we refer the reader to these papers.
There is some overlap between these samples:  3 of the FM cluster sample 
extinction curves and 24 of the FM program sample extinction curves are also in 
the the JG sample.  This lowers the number of Galactic extinction curves in our 
sample from 195 to 168.  Of these extinction curves, Jenniskens \& Greenberg 
(1993) deemed 25 to be of low quality (see Jenniskens \& Greenberg 1993 for 
details).  This lowers the number of Galactic extinction curves in our sample to 
143.

Of the 39 FM program sample lines of sight with high quality extinction curves, 
Cardelli, Clayton, \& Mathis (1989) found values of $R_V$ for 25 of these lines 
of sight from BVRIJHKL photometry.  Of the 90 JG sample lines of sight with high 
quality extinction curves, Aiello \etal (1988) found values of $R_V$ for 49 of 
these lines of sight from BVK photometry.  Eight of these lines of sight are in 
common.  This lowers the number of Galactic values of $R_V$ from 74 to 66.

From the fitted parameter values of the 21 high quality extinction curves that 
the FM and JG samples have in common, Jenniskens \& Greenberg (1993) measured 
systematic and random errors between the two group's fitted values for each 
extinction curve parameter; we list these errors in Table 2.  Furthermore, these 
systematic and random errors are comparable in size.  In the interest of 
creating a uniform data set, primarily to facilitate identification and modeling 
of the correlations between these parameters, we have shifted each group's 
fitted parameter values by one half of the systematic difference between the two 
group's results; this brings both group's results into general agreement.  We 
re-inject these systematic errors into the analysis in \S 4.2.

Secondly, unlike the fitted parameter values of the LMC and SMC extinction 
curves, which we introduce below, uncertainties were not determined for each, or 
any, of the fitted parameter values of the Galactic extinction curves.  
Consequently, we adopt the above measured random errors as indicative of the 
uncertainties in the fitted parameter values of each of the Galactic extinction 
curves in our sample.
Technically, since both groups fitted to the same IUE spectra, these random 
errors are lower limits.  However, Massa \& Fitzpatrick (1986) measured nearly 
identical upper limits from variations in the fitted parameter values of 
extinction curves measured along different lines of sight in the same OB 
associations; we list these errors in Table 2 also.
We adopt the lower limits because this forces us to conservatively overestimate 
the extrinsic scatters in the fits of \S 4.2.

To our Galactic sample, we have added the results of fits to 19 extinction 
curves, and 10 corresponding values of $R_V$, from the LMC (Misselt, Clayton \& 
Gordon 1999), and the results of fits to 4 extinction curves, and 3 
corresponding values of $R_V$, from the SMC (Gordon \& Clayton 1998).  This 
raises the number of extinction curves in our sample to 166, and the number of 
corresponding values of $R_V$ to 79.  Uncertainties have been determined for 
each of the fitted parameter values of the LMC and SMC extinction curves.

\subsection{Correlations between Dust Extinction Curve Parameters}

We now model known correlations between the pairs of extinction curve 
parameters $c_1$ and $c_2$, and $R_V$ and $c_2$.  We also present
possible  correlations between the triplets of extinction curve
parameters $c_3$,  $\gamma$, and $c_2$; $c_4$, $c_3$, and $c_2$; and
$c_4$, $\gamma$, and $c_2$; however, we consider these three-parameter
correlations to be too speculative to incorporate into our extinction
curve prior.  We also constrain the values of the extinction curve
parameters $\gamma$ and $x_0$, which do not vary significantly across
all 166 of the lines of sight in our data set (\S 4.1).  Finally, we
use this information to construct the extinction curve prior.

We begin with a discussion of the extinction curve parameter $c_2$,
upon which all of the above correlations depend.  In \S 3.2, we
pointed out that different values of $c_2$ are measured from different
interstellar environments:  low values of $c_2$ are measured from
young star-forming regions, low to moderate values of $c_2$ are
measured from dense clouds, moderate values of $c_2$ are measured from
the diffuse ISM, and moderate to high values of $c_2$ are measured
from old star-forming regions.  This is not to say that $c_2$ is
a good measure of an interstellar environment's type:  e.g., 30
Doradus is roughly ten times as active of a star-forming region as any
region in the SMC (Caplan et al. 1996), yet significantly higher
values of $c_2$ are measured from the SMC bar than from 30 Doradus; 
this is probably due to the SMC having lower density clouds, which are
less able to protect the classical grains from UV radiation and
supernovae shocks, than does the LMC (Misselt, Clayton, \& Gordon 1999). 
However, $c_2$ might be a reasonable, approximate measure of the net
ability of an interstellar environment to affect extinguishing
grains.  Hence, the above correlations might be viewed as how the
values of other extinction curve parameters, or the correlations
between other extinction curve parameters, vary as a function of a
single-parameter measure of net environmental conditions.

To apply the statistical methodology that we presented in \S 2.2, (1)
the models that we adopt to describe the above correlations must be
slowly varying, i.e., the curve $y = y_c(x;\theta_m)$ from \S 2.2.2 and
\S 2.2.3 must be varying from linearity only on scales that are larger
than the scales given by the scatter of the data that we presented in
\S 4.1 about this curve; and (2) the density of these data along this
curve also must be slowly varying; i.e., the intrinsic density along
this curve, $g(x,y_c)$, and the selection function, $f(x,y)$, must be
varying from constancy only on scales that are larger than the scales
given by this scatter of the data (\S 2.2.2).  As we adopt only
constant, linear, and slowly-varying quadratic models below, the first
condition is met.  As for the second condition, the density of the data
varies most obviously with the value of $c_2$.  This is probably due to
the selection function; e.g., diffuse ISM lines of sight, and
consequently, values of $c_2 \sim 0.8$, have been selected more often
than lines of sight through any other type of interstellar environment,
or value of $c_2$.  These density variations, however, also occur on
scales that are larger than the scales given by the scatter of the
data; consequently, the second condition also appears to be met. 
Hence, we appear to be within the realm of the formalism that we
presented in \S 2.2, with but one caveat.  The intrinsic scatters of
some of the extinction parameters, namely $c_1$ and $c_2$, are probably
somewhat correlated (Fitzpatrick \& Massa 1988).  This causes us to
somewhat underestimate the extrinsic scatters of these correlations
below, but not significantly. 

We begin with the two-parameter correlations:  $c_1$ and $c_2$ (Figure 2), and 
$R_V$ and $c_2$ (Figure 3).  In both cases, the parameters are well correlated; 
however, both of these correlations are more mathematical than physical in 
nature.  In the case of the former correlation, the linear component of Equation 
(\ref{fm}) is observed to pivot about a point at a wavelength of $\lambda 
\approx 3000$ \AA$ $ as the slope, $c_2$, of this component changes from one 
line of sight to another.  If $c_1$ were measured at this wavelength, instead of 
at $\lambda = \infty$, this correlation would disappear; consequently, this 
correlation is merely an artifact of the fitting procedure by which the values 
of these parameters were determined (Carnochan 1986), and not due to any 
intrinsic physical property.
However, the small size of the extrinsic scatter that we measure below for this 
correlation testifies to the constancy of the extinction curve at this pivot 
point.  This is a physical property, since the wavelength of this pivot point 
differs from the V-band wavelength, $\lambda \approx 5500$ \AA, at which the 
extinction curve is normalized.

To the degree that the IR and optical extinction curve is really a one-parameter 
function of $R_V$, and to the degree that the linear component of the UV 
extinction curve is really a one-parameter function of $c_2$ (since $c_1$ and 
$c_2$ are strongly correlated), $R_V$ and $c_2$ must be correlated if the 
extinction curve is to be continuous and differentiable between optical and UV 
wavelengths; 
consequently, this relation is also mathematical in nature.  However, physical 
information can be gleamed from the value of $R_V$, which by Equation 
(\ref{rv2}) is a measure of the relative numbers and/or absorptivities of grains 
extinguishing in the $B$ band to grains extinguishing in the $V$ band.  
How this value changes as the value of $c_2$ is changed provides physical 
information, perhaps as a function of environmental conditions, as we have 
discussed above.

We now model these two correlations, and construct the prior, as described in \S 
2.2.2 and \S 2.2.3.  We do not include the Orion Nebula lines of sight in either 
of these fits, because nebular background contamination artificially lowers the 
measured values of $c_1$ along these lines of sight (Panek 1983), and similarly 
may affect the measured values of $c_2$ (Fitzpatrick \& Massa 1988); this effect 
has not been corrected for in these data (Fitzpatrick \& Massa 1988).

Given how the values of $c_1$ and $c_2$ were determined, we model the first of 
these correlations with a function that is linear in $c_2$:
\begin{equation}
c_1(c_2) = b + m(c_2 - \bar{c_2}),
\end{equation}
where $\bar{c_2}$ is the sample's median value of $c_2$.  
The 1-$\sigma$ uncertainty in $c_1$ as a function of $c_2$ is approximately 
given by
\begin{equation}
\sigma_{c_1}(c_2) = \left[\left(\sigma_b\frac{\partial c_1}{\partial b}\right)^2 
+ \left(\sigma_m\frac{\partial c_1}{\partial m}\right)^2 + 
\left(\sigma_{c_1}\right)^2  + \left(\sigma_{c_2}\frac{\partial c_1}{\partial 
c_2}\right)^2\right]^{1/2},
\label{c1sig}
\end{equation}
where $\sigma_b$ and $\sigma_m$ are the fitted 1-$\sigma$ uncertainties in the 
parameters $b$ and $m$, and $\sigma_{c_1}$ and $\sigma_{c_2}$ are the fitted 
1-$\sigma$ extrinsic scatters in the $c_1$ and $c_2$ dimensions (\S 2.2.3).
Assuming a flat prior, we find:  $\bar{c_2} = 0.711$, $b = -0.064$, $\sigma_b = 
0.026$, $m = -3.275$, $\sigma_m = 0.083$, $\sigma_{c_1} = 0.088$, and 
$\sigma_{c_2} = 0.008$.  Using Equation (\ref{c1sig}) and these fitted values, 
we plot approximate 1-, 2-, and 3-$\sigma$ confidence regions in Figure 2.  
Re-injection of the systematic errors between the FM and JG samples that we 
removed in \S 4.1 (Table 2) increases $\sigma_{c_1}$ to 0.176 and $\sigma_{c_2}$ 
to 0.037. 

The $R_V$-$c_2$ correlation clearly is non-linear; however, it is slowly 
varying, so we model it with a function that is quadratic in $c_2$:
\begin{equation}
R_V(c_2) = b + m(c_2 - \bar{c_2}) + n(c_2 - \bar{c_2})^2.
\end{equation}
The 1-$\sigma$ uncertainty in $R_V$ as a function of $c_2$ is approximately 
given by
\begin{equation}
\sigma_{R_V}(c_2) = \left[\left(\sigma_b\frac{\partial R_V}{\partial b}\right)^2 
+ \left(\sigma_m\frac{\partial R_V}{\partial m}\right)^2 + 
\left(\sigma_n\frac{\partial R_V}{\partial n}\right)^2 + 
\left(\sigma_{R_V}\right)^2  + \left(\sigma_{c_2}\frac{\partial R_V}{\partial 
c_2}\right)^2\right]^{1/2}, 
\label{rvsig}
\end{equation}
where $\sigma_n$ is the fitted 1-$\sigma$ uncertainty in the parameter $n$, and 
$\sigma_{R_V}$ is the fitted 1-$\sigma$ extrinsic scatter in the $R_V$ 
dimension.
Assuming a flat prior, we find:  $\bar{c_2} = 0.721$, $b = 3.228$, $\sigma_b = 
0.053$, $m = -2.685$, $\sigma_m = 0.159$, $n = 1.806$, $\sigma_n = 0.129$, 
$\sigma_{R_V} = 0.000$, and $\sigma_{c_2} = 0.142$.  Using Equation 
(\ref{rvsig}) and these fitted values, we plot approximate 1-, 2-, and 
3-$\sigma$ confidence regions in Figure 3.  Re-injection of the systematic 
errors between the FM and JG samples increases $\sigma_{R_V}$ to 0.112 and 
$\sigma_{c_2}$ to 0.147.

We now consider the three-parameter correlation $c_3$, $\gamma$, and $c_2$ 
(Figure 4).  A strong correlation exists between $c_3$ and $\gamma$ for Galactic 
lines of sight ($c_2 \sim 2/3$) (Fitzpatrick \& Massa 1988; Jenniskens \& 
Greenberg 1993); however, inclusion of the Orion Nebula-like lines of sight 
($c_2 \sim 0$), and the LMC and SMC wing lines of sight ($c_2 \sim 4/3$) ruins 
this previously-determined correlation; the SMC bar lines of sight ($c_2 \sim 
7/3$) are not constraining since $\gamma$ cannot be well determined when $c_3 
\sim 0$.  These lines of sight request a shallower relation.  Physically, this 
probably corresponds to the destruction or alteration of the UV bump grains by 
UV radiation and/or supernovae shocks (\S 3.2, Figure 5; see also Clayton, 
Gordon, \& Wolff 2000).

Weak correlations exist between $c_4$ and $c_3$, and $c_4$ and $\gamma$ for 
Galactic lines of sight (Fitzpatrick \& Massa 1988; Jenniskens \& Greenberg 
1993); indeed, weak positive correlations can be seen in Figure 6, if only the 
Galactic lines of sight are considered.
However, inclusion of the LMC and SMC lines of sight also ruins these 
previously-determined correlations.  The destruction or alteration of UV bump 
grains by the environment probably accounts for the shift to lower values of 
$c_3$ in the top panel of Figure 6.  Hydrogenation might account for the greater 
scatter in the bottom panel of Figure 6 (\S 3.2).

In any case, we do not attempt to model and constrain the possible correlations 
in Figures 4, 5, and 6.  First of all, in distant galaxies, these grain species 
might occur in different relative abundances, perhaps due to different relative 
metallicities.  Secondly, even if this is not the case, if the extinction is due 
primarily to dust that is local to the burst, the relative abundances of these 
grain species may be altered by the burst itself, as well as by the afterglow 
(Lamb \& Reichart 2000b).  Consequently, no constraint can be placed between the 
extinction curve parameters $c_2$, $c_3/\gamma^2$, and $c_4$.

The values of $\gamma$ and $x_0$ are approximately constant across all 166 lines 
of sight.  We find their values to be $\gamma = 0.958 \pm 0.088$ and $x_0 = 
4.593 \pm 0.020$.  Since the width of the UV bump is approximately equal to the 
widths of the photometric bands ($\Delta \nu / \nu \approx 0.2$), and since the 
uncertainty in the center of the UV bump is significantly smaller than the 
widths of the photometric bands ($\Delta \nu / \nu \approx 0.004 \ll 0.2$), when 
fitting to afterglow photometry, precise values of these parameters cannot be 
extractable from the data; i.e., the fitted solutions should largely resemble 
the adopted prior, particularly in the case of the bump center parameter, $x_0$.

We now construct from these results an extinction curve prior, in accordance 
with the examples of \S 2.2.1.  For the extinction curve parameters $R_V$, 
$c_1$, $c_2$, $\gamma$, and $x_0$, we recommend that the following prior be 
used:
\begin{eqnarray}
\nonumber
p(R_V,c_1,c_2,\gamma,x_0|I) = 
G[R_V,R_V(c_2),3\sigma_{R_V}(c_2)]G[c_1,c_1(c_2),3\sigma_{c_1}(c_2)]\\
\times G(\gamma,0.958,0.264)G(x_0,4.593,0.060).
\end{eqnarray}
Here, we have conservatively tripled the 1-$\sigma$ uncertainties of the 
component priors, simply because these priors are determined solely from 
information that is local to our galaxy.  For the extinction curve parameters 
$A_V$, $c_3/\gamma^2$, and $c_4$, we conservatively recommend that a flat prior 
be used.  Altogether, this prior weights the eight-dimensional parameter space 
of the extinction curve model that we presented in \S 3 such that the volume of 
the solution space is reduced significantly, {\it a priori}.  

Finally, we comment on the possibility of an evolving extinction curve.  As 
mentioned above, if an afterglow is extinguished by dust that is local to a 
burst, energetic photons, both from the burst and from the afterglow, may alter 
the extinction curve with time (Lamb \& Reichart 2000b).  However, since all 
afterglows observed to date have faded more rapidly than $F_{\nu} \sim t^{-1}$ 
at optical through X-ray wavelengths, the majority of these energetic photons 
are probably emitted during the first few seconds or minutes of the afterglow, 
if not during the burst itself.
Hence, any dust destruction or alteration that may occur, should occur on such 
timescales.  By restricting oneself to photometry taken hours or longer after a 
burst, one should be able to safely ignore the possibility of an evolving 
extinction curve.

\section{The Ly$\alpha$ Forest Flux Deficit Model and Prior}

At redshifts of $z \ga 2$, the Ly$\alpha$ forest will absorb light at optical 
wavelengths, and consequently cannot be ignored (Fruchter 1999a; Lamb \& 
Reichart 2000a).  We present a two-parameter model that describes the effects of 
the Ly$\alpha$ forest on the spectral flux distributions of afterglows (and on 
the spectral flux distributions of all extragalactic point sources for that 
matter).  Using Ly$\alpha$ forest flux deficit measurements from quasar 
absorption line systems, we construct a prior that weights this two-dimensional 
space such that the volume of the solution space is reduced significantly. 
 
In the study of quasar absorption line systems, the quantity called flux 
deficit, denoted $D_A$, is defined by
\begin{equation}
D_A = \left<1-\frac{F_\nu({\rm observed})}{F_\nu({\rm continuum})}\right>,
\end{equation}
where this quantity is averaged over the wavelength range between the emission 
lines Ly$\alpha$ and Ly$\beta$ $+$ O VI that is not affected by emission line 
wings, and only if the continuum can be reliably extrapolated from the 
unabsorbed spectrum at longer wavelengths (Oke \& Korycansky 1982).  Zuo \& 
Phinney (1993), Zuo (1993), and Lu \& Zuo (1994) model this quantity by
\begin{equation}
D_A(z) \approx 1 - \exp\left[-a\left(\frac{1+z}{1+\bar{z}}\right)^b\right],
\end{equation}
where $a$ and $b$ are parameters whose values are determined by fitting to flux 
deficit measurements, $z$ is the redshift corresponding to the central 
wavelength of the range over which the quantity $D_A$ is averaged, and $\bar{z}$ 
is the median value of $z$ for the sample to which one is fitting.
The 1-$\sigma$ uncertainty in $D_A$ as a function of $z$ is approximately given 
by 
\begin{equation}
\sigma_{D_A}(z) = \left[\left(\sigma_a\frac{\partial D_A}{\partial a}\right)^2 + 
\left(\sigma_b\frac{\partial D_A}{\partial b}\right)^2 + 
\left(\sigma_{D_A}\right)^2  + \left(\sigma_{z}\frac{\partial D_A}{\partial 
z}\right)^2\right]^{1/2},
\label{dasig}
\end{equation}
where $\sigma_a$ and $\sigma_b$ are the fitted 1-$\sigma$ uncertainties in the 
parameters $a$ and $b$, and $\sigma_{D_A}$ and $\sigma_z$ are the fitted 
1-$\sigma$ extrinsic scatters in the $D_A$ and $z$ dimensions (\S 2.2.3).

Assuming a flat prior, and adopting Sample 4 of Zuo \& Lu (1993), which is a 
combination of Sample 2 of Zuo \& Lu (1993) (see Zuo \& Lu 1993 for details) and 
the high redshift ($z \sim 4$) sample of Schneider, Schmidt, \& Gunn (1989a,b, 
1991), we find:  $\bar{z} = 2.994$, $a = 0.306$, $\sigma_a = 0.010$, $b = 
4.854$, $\sigma_b = 0.188$, $\sigma_{D_A} = 0.000$, and $\sigma_z = 0.165$.
We plot Sample 4 of Zuo and Lu (1993) and, using Equation (\ref{dasig}) and 
these fitted values, approximate 1-, 2-, and 3-$\sigma$ confidence regions in 
Figure 7.  The extent of the scatter about the best fit in Figure 7 is largely a 
reflection of the extent of the wavelength range over which these values of 
$D_A$ were averaged.  This wavelength range corresponds to $\Delta \nu / \nu 
\approx 0.2$, which is typical of the photometric bands.  In other words, the 
scatter in Figure 7, very conveniently, is typical of what one would find if 
Ly$\alpha$ forest flux deficits were measured from afterglows photometrically.  
Consequently, the Ly$\alpha$ forest flux deficit prior {\it for photometric, as 
opposed to spectroscopic, data} is given by (\S 2.2.1)
\begin{equation}
p(D_A,z|I) = G[D_A,D_A(z),\sigma_{D_A}(z)]. 
\end{equation}

\section{Example Extinguished and Absorbed Spectral Flux Distributions}

We now demonstrate the breath of the models of \S 3 and \S 5.  Using
these models, we plot in Figure 8 example spectral flux distributions
that have been extinguished by dust in a host galaxy, absorbed by H I in the 
host galaxy, redshifted, and absorbed by the Ly$\alpha$ forest, for a
wide variety of plausible extinction curves and redshifts.  We have
adopted an intrinsic spectrum of $F_\nu \propto \nu^{-1}$, and we allow
the values of $A_V$, $c_2$, $c_3/\gamma^2$, $c_4$, and $z$ to vary over
observed/reasonable ranges.  The values of $R_V$, $c_1$, $\gamma$,
$x_0$, and $D_A$, we take from the best fits of \S 4.2 and \S 5.  
Finally, we convolve each extinguished, absorbed, and redshifted spectrum
with a logarithmically flat smearing function of width $\Delta \nu =
0.2\nu$, converting each spectrum to a spectral flux distribution;
i.e., we model how these spectra would be sampled photometrically, as
opposed to spectroscopically (\S 5).  Clearly, a single intrinsic
spectrum can manifest itself in a multitude of ways, and exhibit a
variety of broad spectral features, including a shoulder in the
infrared, the UV bump, the Ly$\alpha$ forest, and the Lyman limit.

\section{Conclusions}

In this paper, we have presented a very general, Bayesian inference formalism 
with which afterglow models can be tested, and with which the parameter values 
of acceptable models can be constrained.  Furthermore, we have developed and 
presented a formalism for the construction of Bayesian prior probability 
distributions from multi-dimensional data sets, which we have drawn on 
extensively.  We have presented models that describe how extinction by dust, 
both in host galaxies and in our galaxy, and absorption by the Ly$\alpha$ forest 
and by H I in host galaxies, change the intrinsic spectra of afterglows.  Then, 
applying the above formalism, we constructed a prior that weights the 
additional, but necessary, parameter space of these models such that the volume 
of the solution space is reduced significantly, {\it a priori}.  These models 
and priors will lead to the more realistic modeling of afterglows, particularly 
at IR through UV wavelengths, in future papers.

Finally, we emphasize that the phenomena for which we have presented models and
priors in this paper -- extinction by dust and absorption by the Ly$\alpha$
forest -- affect identically the light from all other extragalactic point
sources.\footnote{By point source, we mean either that the host galaxy
contributes a negligible fraction of the total light within the point spread
function of the point source, or that this contribution of the host galaxy to
the total light can be measured directly -- which can be done in the case of a
fading point source after it fades away -- and consequently separated from that
of the point source.  Otherwise, one must model not a single point source in a
distribution of dust, but instead a distribution of point sources in a
distribution of dust, which is a significantly more challenging endeavor, but
certainly not impossible (e.g., Witt, Thronson, \& Capuano 1992; Gordon,
Calzetti, \& Witt 1997).  Similarly, light from a point source that is either
scattered or absorbed and thermally re-emitted into the line of sight can
contribute non-negligibly to, and even dominate the direct light of a fading
point source at late times, due to the time delay with which the indirect light
is received (e.g., Reichart 2000).  Again, one must either use data from early
times, when the contribution of the ``dust echo'' is negligible, or measure the
contribution of the dust echo at late times, when the contribution of the fading
point source is negligible, and use this information to properly interpret the
data at intermediate times, when neither component can be ignored.}
Consequently, the work presented in this paper is as applicable to high-redshift
Type Ia supernovae and quasars, for example, as it is to the afterglows of
bursts.  Since the effects of extinction and absorption are most dramatic at UV
wavelengths in the source frame, these models and priors will be particularly
useful for the modeling of optical photometry of high-redshift point sources.

\acknowledgements

Support for this work was provided by NASA through the Hubble Fellowship grant
\#HST-SF-01133.01-A from the Space Telescope Science Institute, which is
operated by the Association of Universities for Research in Astronomy, Inc.,
under NASA contract NAS5-26555.  Support for this work was also provided by NASA 
contracts NASW-4690 and SCSV 464006.  I am very grateful to C. Graziani, D. Q. 
Lamb, and M. C. Miller, in particular for many discussions that we had 
concerning the Bayesian inference techniques presented in this paper.  I am also 
grateful to S. Burles for discussions that we had concerning the modeling of the 
Ly$\alpha$ forest flux deficit measurements, and to A. N. Witt for discussions 
that we had concerning the physical nature of the grain species.

\clearpage

\begin{deluxetable}{ccccc}
\footnotesize
\tablecolumns{5}
\tablewidth{0pc}
\tablecaption{Breakdown of the Extinction Curve Data Set}
\tablehead{\colhead{Galaxy} & \colhead{Extinction Curve  
Sample\tablenotemark{a}} & \colhead{Number of Extinction Curves} & 
\colhead{Number of Values of $R_V$} & \colhead{$R_V$ 
Reference\tablenotemark{a}}}
\startdata
MW & FM Cluster & 35 & 0 & \nl
 & FM Program & 45 & 25 & CCM \nl
 & JG & 115 & 49 & Aea \nl
 & Combined\tablenotemark{b} & 143 & 66 & \nl
LMC & MCG & 19 & 10 & MCG \nl
SMC & GC & 4 & 3 & GC \nl
Combined & Combined & 166 & 79 & \nl
\enddata
\tablenotetext{a}{FM -- Fitzpatrick \& Massa 1990; JG -- Jenniskens \& Greenberg 
1993; MCG -- Misselt, Clayton, \& Gordon 1999; GC -- Gordon \& Clayton 1998; CCM 
-- Cardelli, Clayton, \& Mathis 1989; Aea -- Aiello et al. 1988.}
\tablenotetext{b}{Overlap between the samples, and low quality data have been 
removed (see \S 4.1 for details).}
\end{deluxetable}

\clearpage

\begin{deluxetable}{cccc}
\footnotesize
\tablecolumns{4}
\tablewidth{0pc}
\tablecaption{Systematic and Random Errors Between the FM and JG Samples}
\tablehead{\colhead{Parameter} & \colhead{Systematic Error\tablenotemark{a,b}} & 
\colhead{Random Error Lower Limit\tablenotemark{b}} & \colhead{Random Error 
Upper Limit\tablenotemark{c}}}
\startdata
$c_1$ & 0.304 & 0.259 & \nl
$c_2$ & -0.073 & 0.050 & 0.08 \nl
$c_3$ & 0.316 & 0.252 & 0.27 \nl
$c_4$ & 0.082 & 0.063 & 0.08 \nl
$\gamma$ & 0.036 & 0.039 & 0.04 \nl
$x_0$ & -0.013 & 0.010 & 0.01 \nl
$R_V$ & -0.224 & 0.205 & \nl
\enddata
\tablenotetext{a}{JG $-$ FM}
\tablenotetext{b}{Based on the 20 non-Orion Nebula, high quality extinction 
curves that the FM and JG samples have in common (see \S 4.1 for details).}
\tablenotetext{c}{Based on variations along different lines of sight in the same 
OB associations (see \S 4.1 for details). From Table 2 of Jenniskens \& 
Greenberg 1993.}
\end{deluxetable}

\clearpage

\figcaption[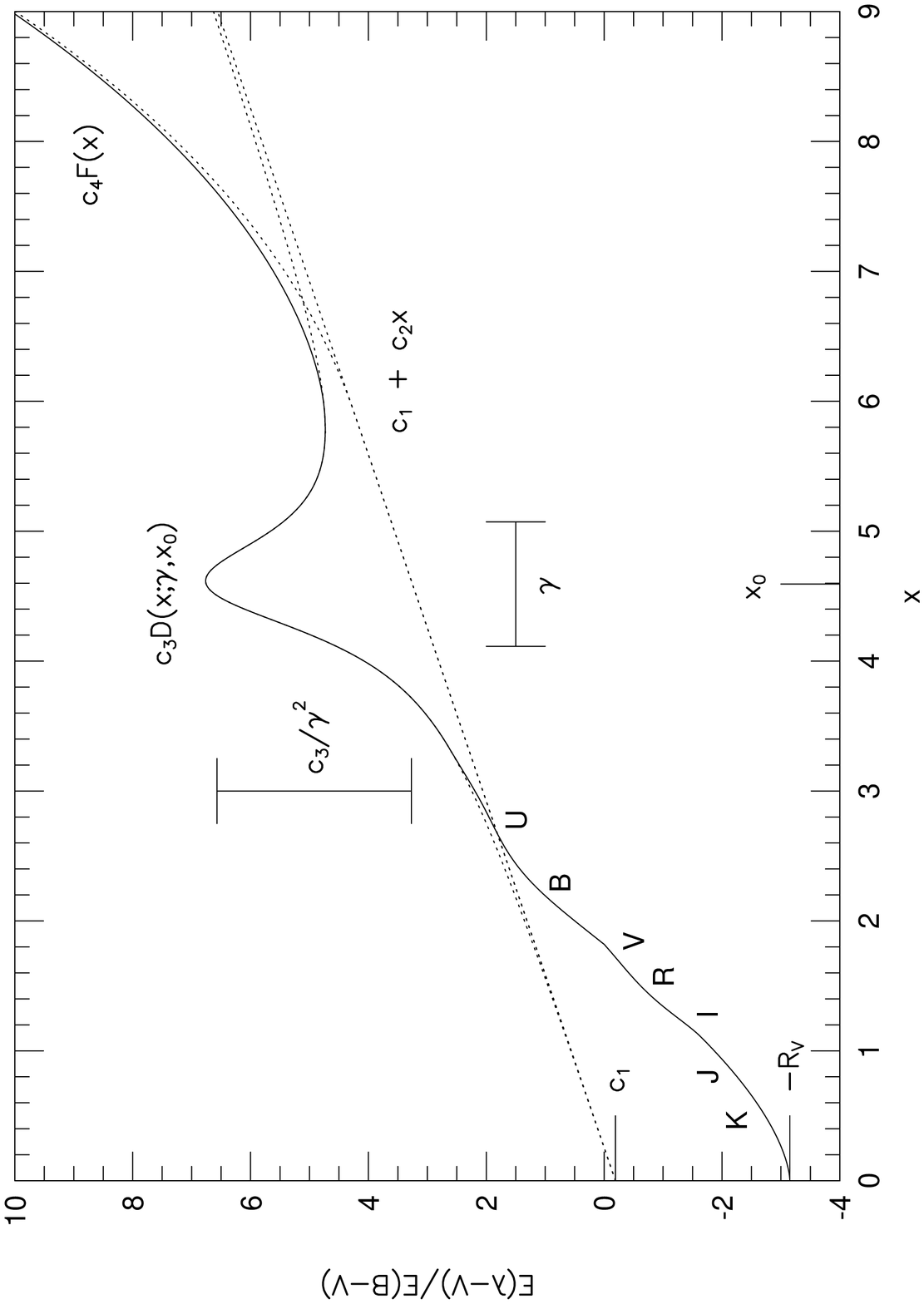]{An extinction curve that is typical of the diffuse ISM of 
our galaxy.  The dotted lines mark the three components of the UV extinction 
curve of Fitzpatrick \& Massa (1988) (see \S 3.2).\label{ext1.ps}}

\figcaption[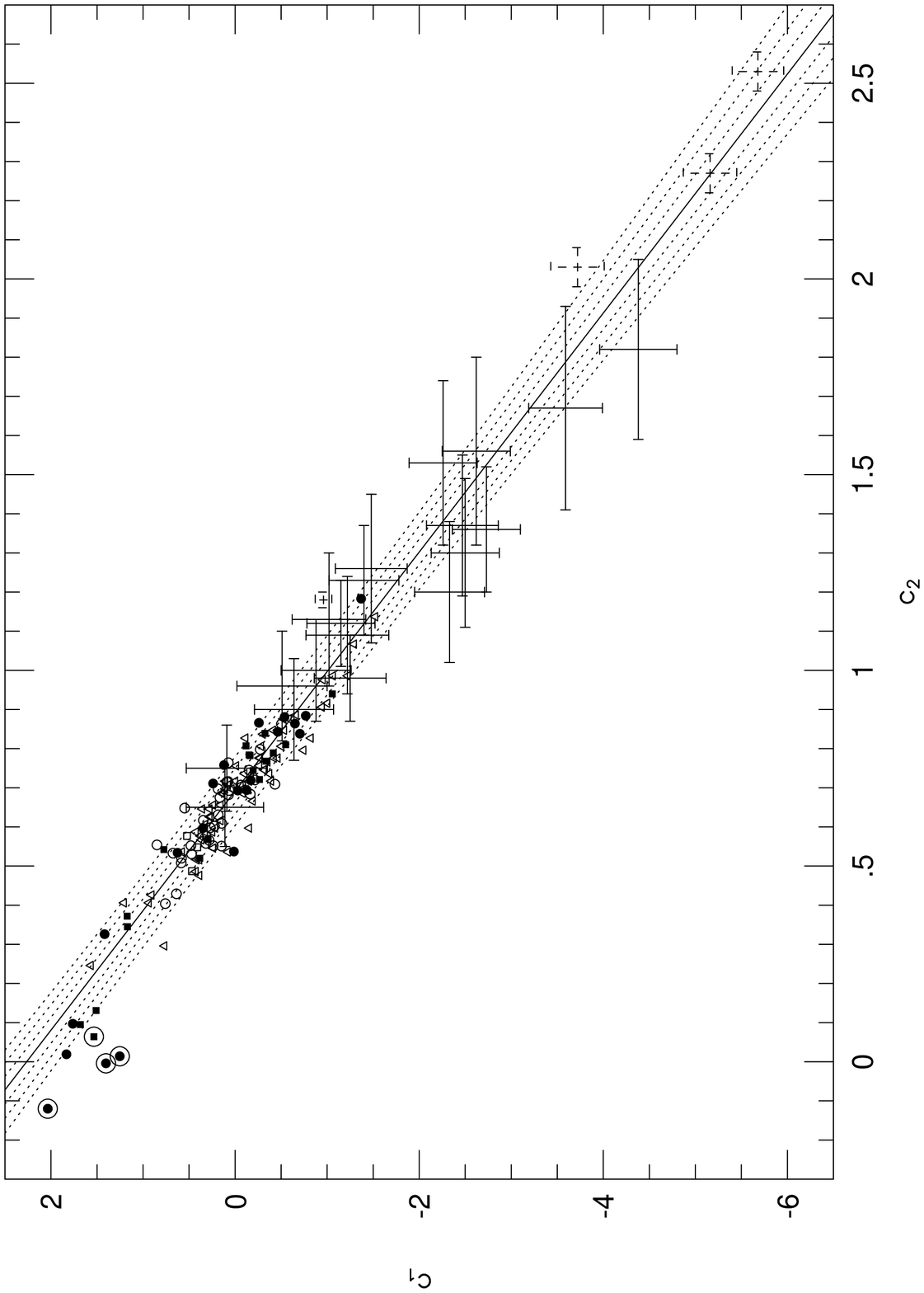]{The correlation between the extinction curve parameters 
$c_1$ and $c_2$.  The dotted lines mark approximate 1-, 2- and 3-$\sigma$ 
confidence regions (see \S 4.2).
The filled squares are from the Fitzpatrick \& Massa (1988; FM) cluster sample, 
the filled circles are from the FM program sample, the empty triangles are from 
the Jenniskens \& Greenberg (1993; JG) sample, the empty squares are from both 
the FM cluster sample and the JG sample, the empty circles are from both the FM 
program sample and the JG sample, the solid error bars denote the Misselt, 
Clayton, \& Gordon (1999) LMC sample, and the dotted error bars denote the 
Gordon \& Clayton (1998) SMC sample (see \S 4.1).  The error bars of the 
Galactic points are discussed in \S 4.1.  The encircled points denote lines of 
sight through the Orion Nebula region.\label{ext2.ps}}

\figcaption[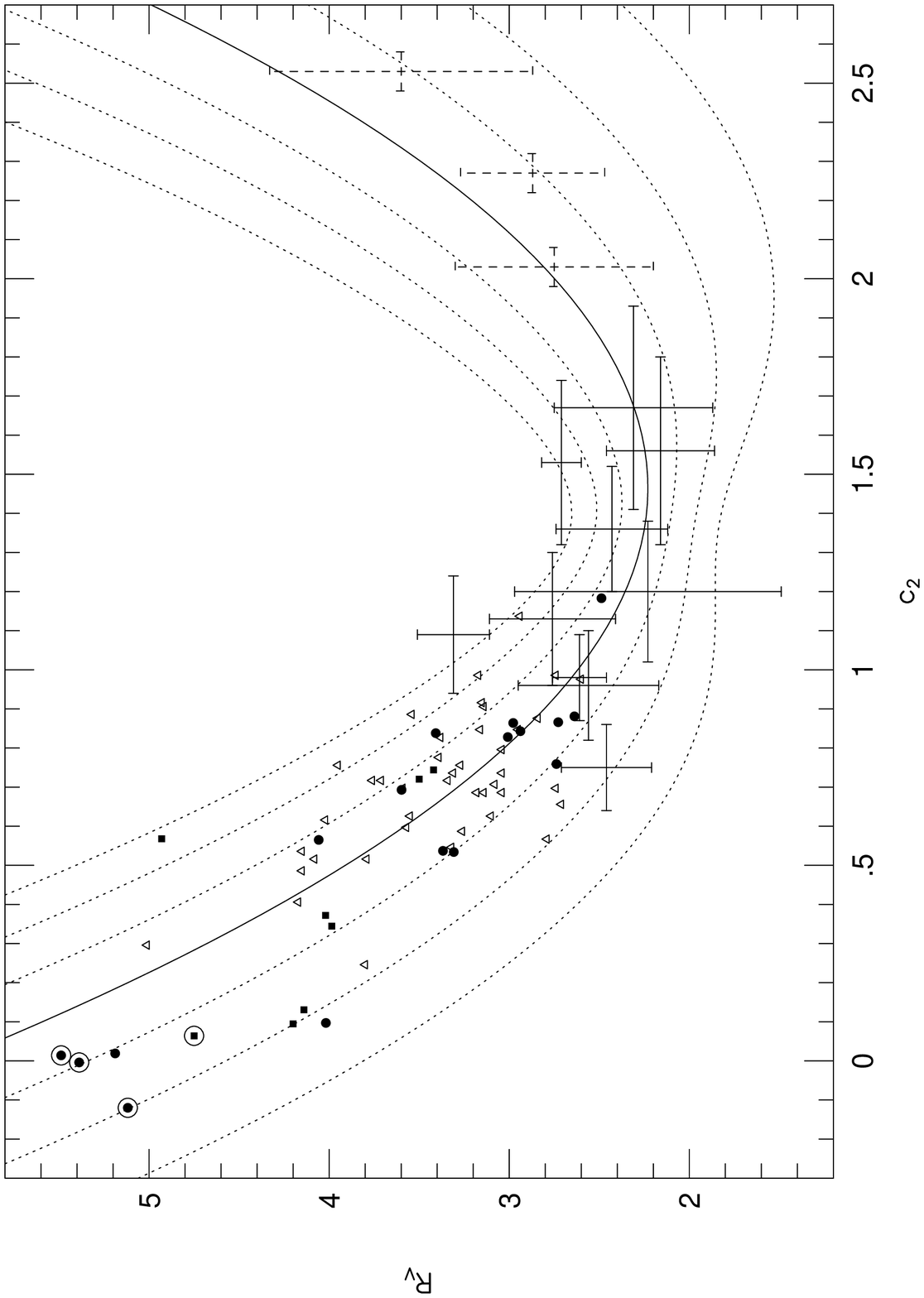]{The correlation between the extinction curve parameters 
$R_V$ and $c_2$.  The dotted lines mark approximate 1-, 2- and 3-$\sigma$ 
confidence regions (see \S 4.2, Figure 2).\label{ext3.ps}}

\figcaption[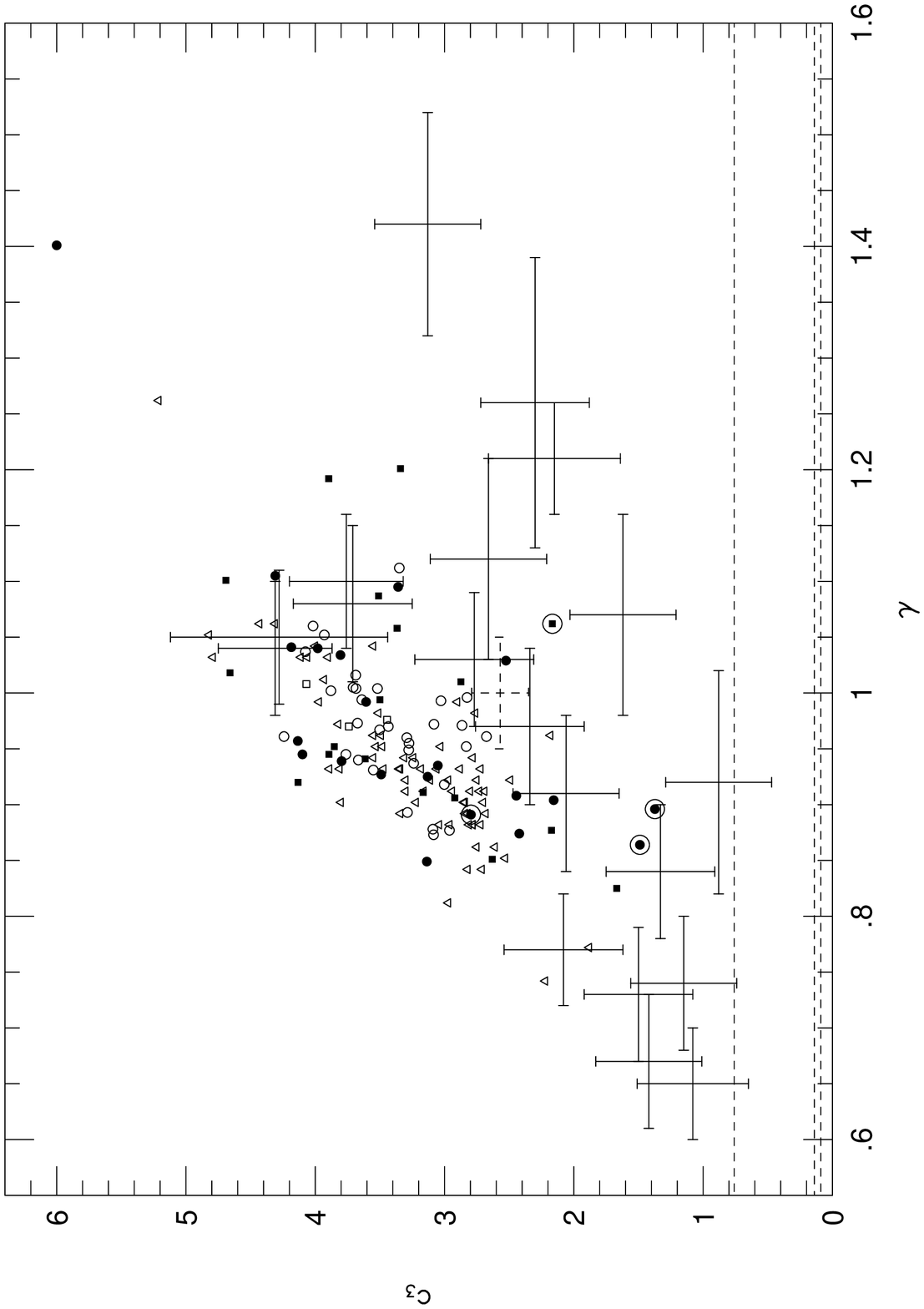]{The correlation between the extinction curve parameters 
$c_3$ and $\gamma$ as a function of $c_2$:  the Orion Nebula region lines of 
sight have $c_2 \sim 0$, the Galactic lines of sight have $c_2 \sim 2/3$, the 
LMC and SMC wing lines of sight have $c_2 \sim 4/3$, and the SMC bar lines of 
sight have $c_2 \sim 7/3$ (see \S 4.2, Figure 2).\label{ext4.ps}}

\figcaption[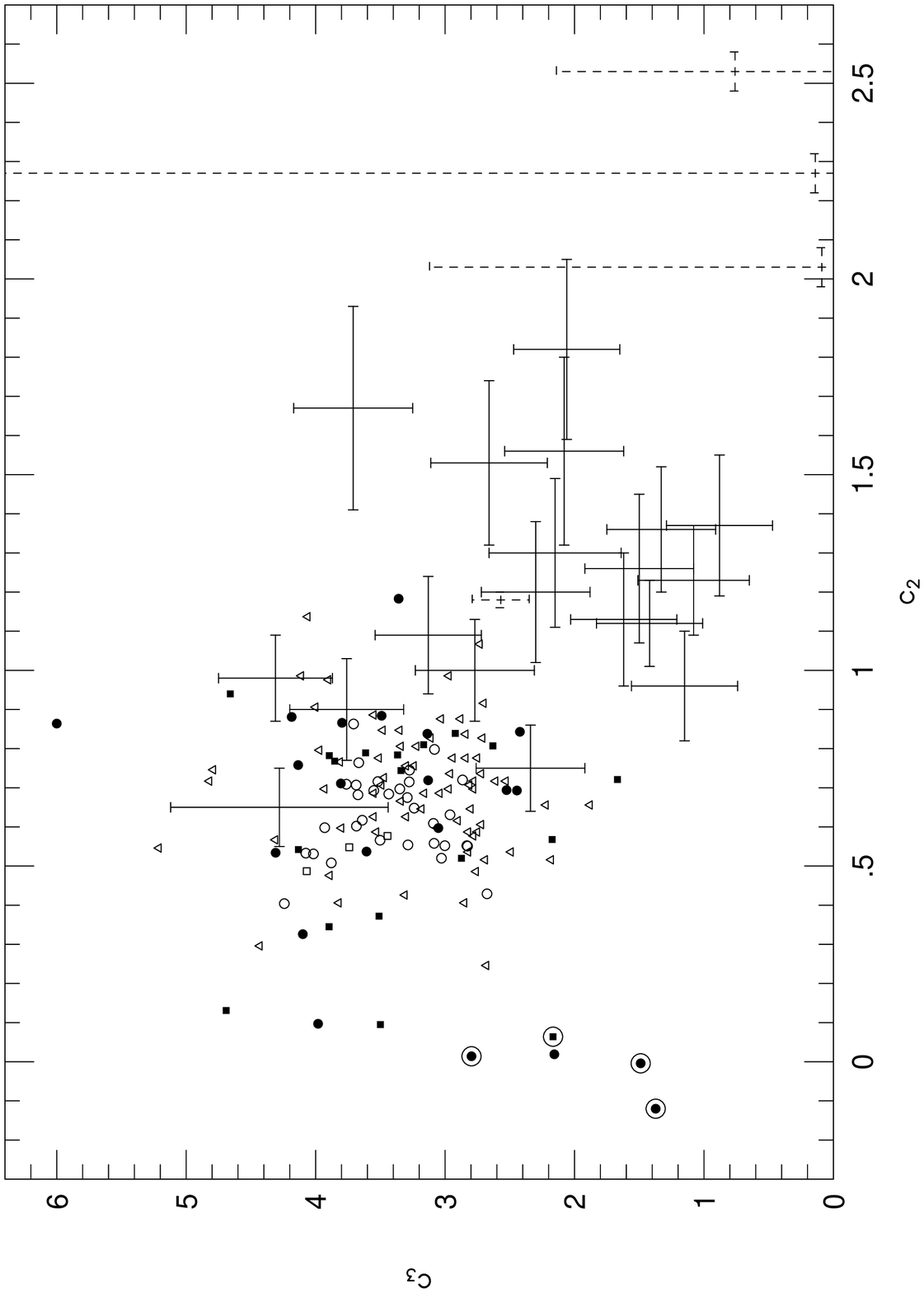]{How the strength of the UV bump, as measured by $c_3$, 
varies with environmental conditions, as measured by $c_2$ (see \S 3.2, \S 4.2, 
Figure 2).\label{ext5.ps}}

\figcaption[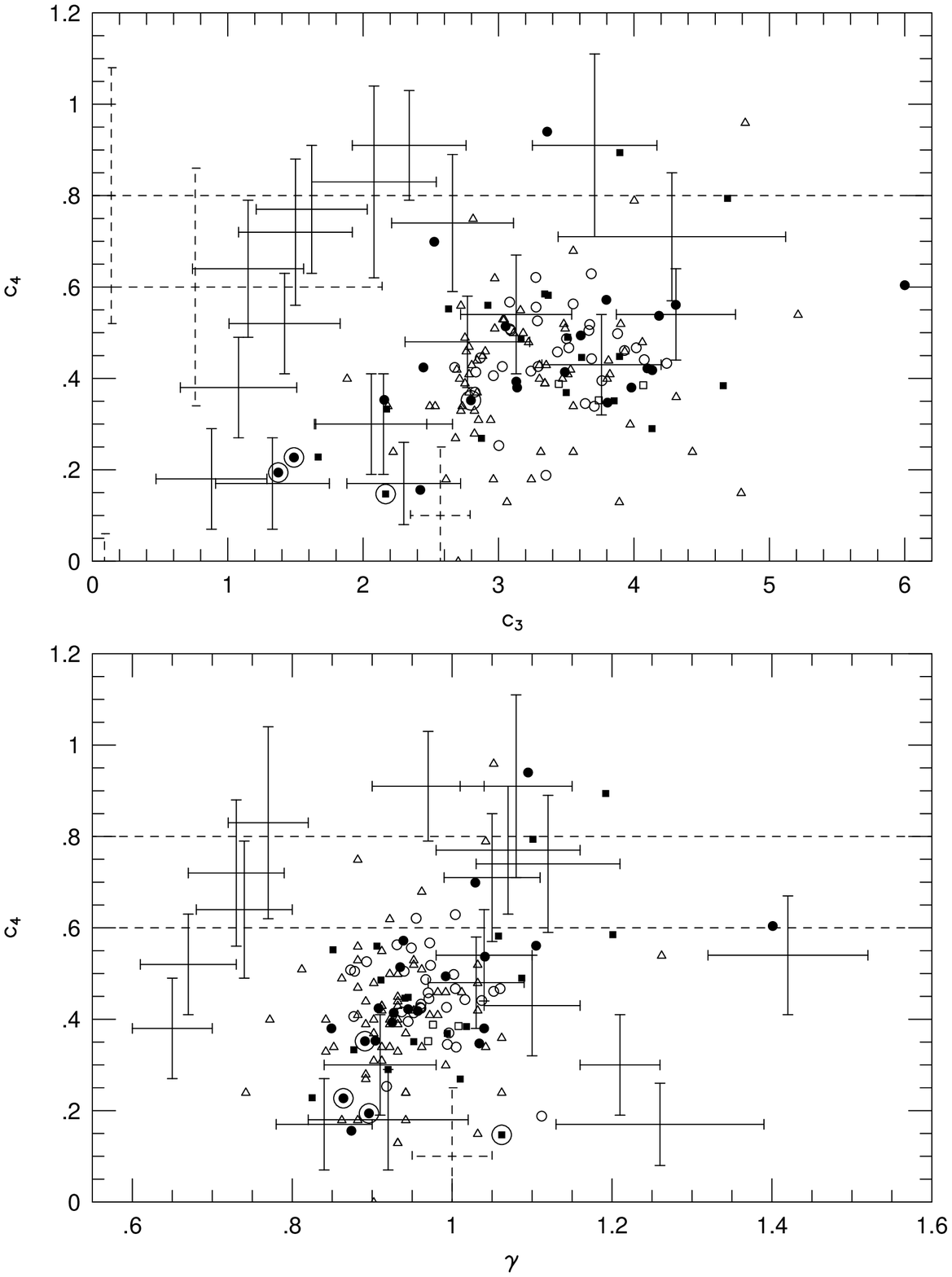]{The correlation between the extinction curve parameters 
$c_4$ and $c_3$ (top panel), and $c_4$ and $\gamma$ (bottom panel) as a function 
of $c_2$ (see \S 4.2, Figure 2, Figure 4).\label{ext6.ps}}

\figcaption[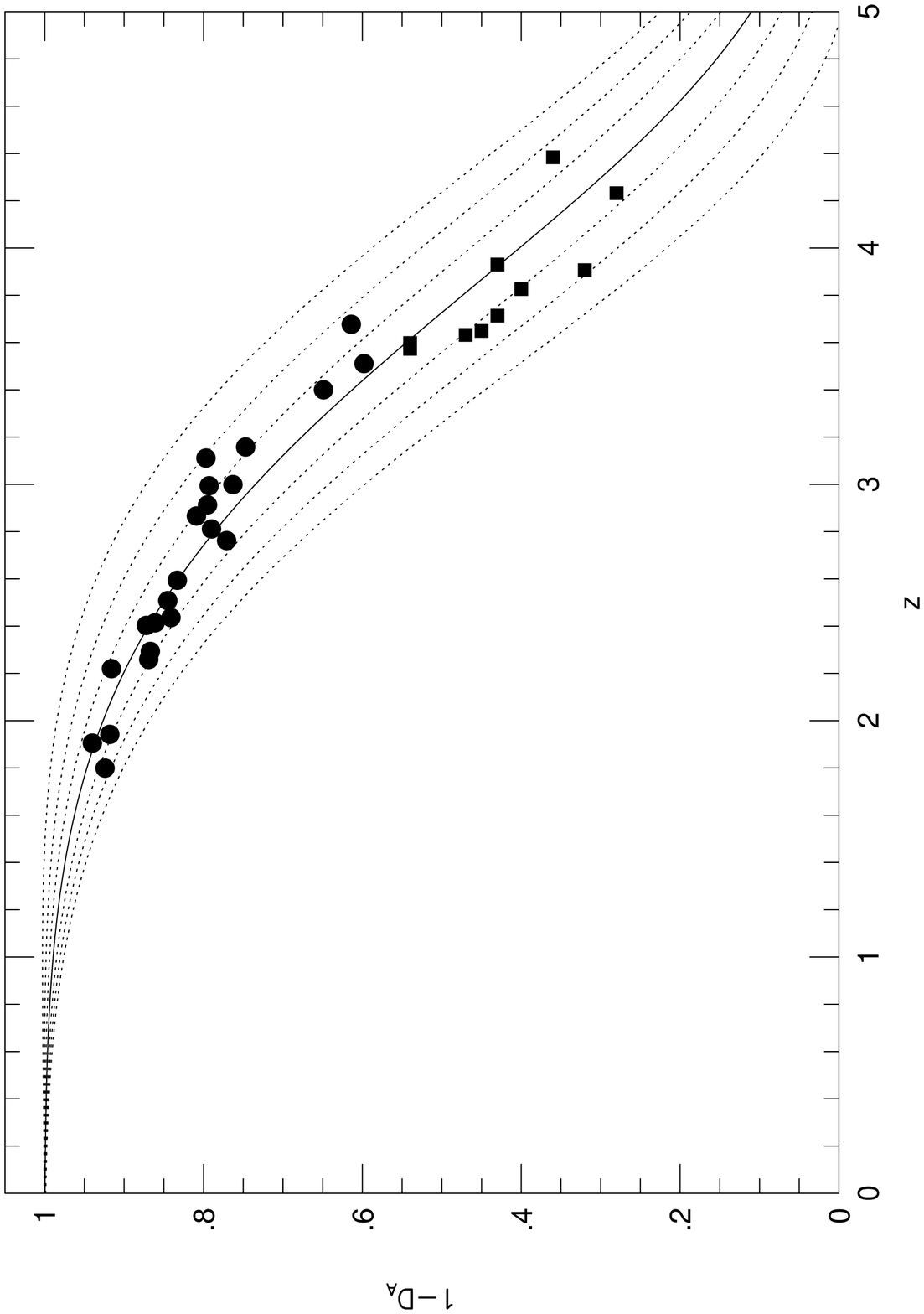]{The correlation between Ly$\alpha$ forest flux deficit, 
$D_A$, and redshift, $z$.  The dotted lines mark approximate 1-, 2- an 
3-$\sigma$ confidence regions (see \S 5).  The circles denote Sample 2 of Zuo \& 
Lu (1993), and the squares denote the high redshift sample of Schneider, 
Schmidt, \& Gunn (1989a,b, 1991).\label{ext7.ps}}

\figcaption[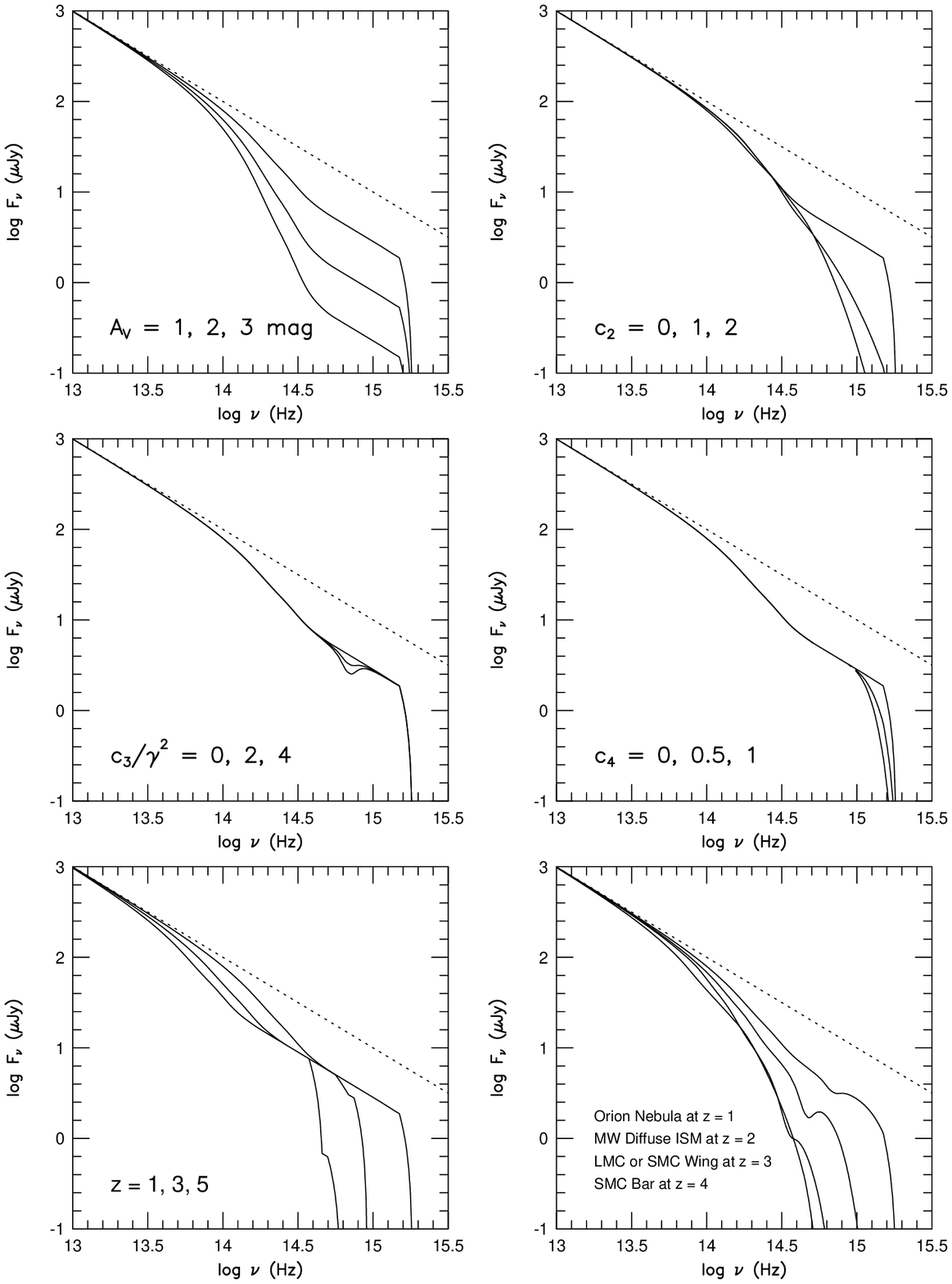]{Example extinguished and absorbed spectral flux 
distributions (see \S 6).  The dotted curve in each panel corresponds to an 
unextinguished, unabsorbed spectral flux distribution, given by $F_\nu \propto 
\nu^{-1}$.  The top solid curve in each of the first five panels is given by 
$(A_V,c_2,c_3/\gamma^2,c_4,z)$ $=$ $(1,0,0,0,1)$.  The lower two solid curves in 
each of these five panels is given by increasing, as marked, the value of a 
single of these five parameters.  In the fifth (redshift) panel, we have fixed 
the spectral flux at long wavelengths.  The solid curves in the sixth panel are 
typical of extinction by dust in (from top to bottom) the Orion Nebula, the 
diffuse ISM of our galaxy, the LMC and the SMC wing, and the SMC bar, for a 
variety of redshifts.\label{ext8.ps}}

\clearpage

\setcounter{figure}{0}

\begin{figure}[tb]
\plotone{ext1.ps}
\end{figure}

\begin{figure}[tb]
\plotone{ext2.ps}
\end{figure}

\begin{figure}[tb]
\plotone{ext3.ps}
\end{figure}

\begin{figure}[tb]
\plotone{ext4.ps}
\end{figure}

\begin{figure}[tb]
\plotone{ext5.ps}
\end{figure}

\begin{figure}[tb]
\plotone{ext6.ps}
\end{figure}

\begin{figure}[tb]
\plotone{ext7.ps}
\end{figure}

\begin{figure}[tb]
\plotone{ext8.ps}
\end{figure}

\end{document}